\documentclass[aps,prb,twocolumn,superscriptaddress,footinbib,amsmath,amssymb]{revtex4}
\usepackage{gensymb}
\usepackage{dcolumn}
\usepackage{bm}
\usepackage[pdftex]{graphicx,color} 
\usepackage{epstopdf} 
\usepackage{graphicx}
\usepackage{float}
\usepackage[english]{babel}
\usepackage[normalem]{ulem} 

\usepackage{array}
\newcolumntype{P}[1]{>{\centering\arraybackslash}p{#1}}
\newcolumntype{M}[1]{>{\centering\arraybackslash}m{#1}}

\begin{document}
\title{Reflectivity calculated for a 3D silicon photonic band gap crystal with finite support}

\author{D. Devashish}
\email{d.devashish@utwente.nl}
\affiliation{Complex Photonic Systems (COPS), 
MESA+ Institute for Nanotechnology, 
University of Twente, P.O. Box 217, 7500 AE Enschede, The Netherlands}
\affiliation{Mathematics of Computational Science (MACS), MESA+ Institute for Nanotechnology, University of Twente, P.O. Box 217, 7500 AE, Enschede, The Netherlands}

\author{Shakeeb B. Hasan}
\email{s.b.hasan@utwente.nl}
\affiliation{Complex Photonic Systems (COPS), 
MESA+ Institute for Nanotechnology, 
University of Twente, P.O. Box 217, 7500 AE Enschede, The Netherlands}

\author{J. J. W. van der Vegt}
\email{j.j.w.vandervegt@utwente.nl}
\affiliation{Mathematics of Computational Science (MACS), MESA+ Institute for Nanotechnology, University of Twente, P.O. Box 217, 7500 AE, Enschede, The Netherlands}

\author{Willem L. Vos}
\email{w.l.vos@utwente.nl}
\affiliation{Complex Photonic Systems (COPS), 
MESA+ Institute for Nanotechnology, 
University of Twente, P.O. Box 217, 7500 AE Enschede, The Netherlands}
\homepage{www.photonicbandgaps.com}


\email{$^*$d.devashish@utwente.nl, URL: www.photonicbandgaps.com} 


\date{September 5, 2016}

\begin{abstract}
We study numerically the reflectivity of three-dimensional (3D) photonic crystals with a complete 3D photonic band gap, with the aim to interpret recent experiments. 
We employ the finite element method to study crystals with the cubic diamond-like inverse woodpile structure.
The high-index backbone has a dielectric function similar to silicon. 
We study crystals with a range of thicknesses up to ten unit cells ($L \leq 10 c$). 
The crystals are surrounded by vacuum, and have a finite support as in experiments. 
The polarization-resolved reflectivity spectra reveal Fabry-P{\'e}rot fringes related to standing waves in the finite crystal, as well as broad stop bands with nearly $100~\%$ reflectivity, even for thin crystals. 
From the strong reflectivity peaks, it is inferred that the maximum reflectivity observed in experiments is not limited by finite size. 
The frequency ranges of the stop bands are in excellent agreement with stop gaps in the photonic band structure, that pertain to infinite and perfect crystals. 
The frequency ranges of the observed stop bands hardly change with angle of incidence, which is plausible since the stop bands are part of the 3D band gap. 
Moreover, this result supports the previous assertion that intense reflection peaks measured with a large numerical aperture provide a faithful signature of the 3D photonic band gap. 
The Bragg attenuation lengths $L_{B}$ exceed the earlier estimates based on the width of the stop band by a factor $6$ to $9$.
Hence crystals with a thickness of $12$ unit cells studied in experiments are in the thick crystal limit ($L >> L_{B}$). 
In our calculations for p-polarized waves, we also observe an intriguing hybridization of the zero reflection of Fabry-P{\'e}rot fringes and the Brewster angle, which has not yet been observed in experiments. 
\end{abstract}

\pacs{42.70.Qs, 42.50.Ct, 88.40.hj}

\maketitle

\section{Introduction}
Numerous efforts are being devoted to the study of the intricate class of three-dimensional (3D) metamaterials known as photonic crystals, that radically control propagation and emission of light.~\cite{Yablonovitch1987PRL,John1987PRL,Joannopoulos1997Nature,Lopez2003AdvMater,Lourtioz2006Book,Joannopoulos2008Book,Ghulinyan2015Book}
These metamaterials are pursued to control spontaneous emission of embedded quantum emitters~\cite{Koenderink2002PRL,Ogawa2004Science,Lodahl2004Nature,Leistikow2011PRL} and cavity quantum electrodynamics (QED)~\cite{Vos2015Chapter}, to control thermal emission~\cite{Fleming2002Nature,Han2007PRL}, to realize efficient miniature lasers~\cite{Tandaechanurat2011NatPhoton},
to efficiently photoelectrically convert light in solar cells ~\cite{Zhou2008JApplPhys}, and for cloaking.~\cite{Ergin2010Science} 
Photonic crystals are composite optical materials in which the refractive index varies spatially with a periodicity on length scales comparable to the wavelength of light. 
Due to the long-range periodic order in photonic crystals, the photon dispersion relations are organized in bands, in analogy to electron bands in a semiconductor.~\cite{Ashcroft1976Book} 
A stop gap is defined as a frequency range for which light cannot travel in a given direction. 
The emergence of a complete 3D photonic band gap, that is, a frequency range for which light is forbidden for all wave vectors and all polarizations is of prime significance to 3D photonic crystals.~\cite{Yablonovitch1987PRL,John1987PRL,Joannopoulos2008Book} 
Since light with frequencies in the photonic band gap cannot exist in the photonic crystal it provides a perfect back reflector in a solar cell for light from any angle and polarization, and enhance the distance light travels inside the solar cell through internal reflections.~\cite{Bermel2007OptExpress} 

The experimental demonstration of a 3D photonic band gap remains a major challenge. 
By definition, a 3D band gap corresponds to a frequency range where the density of optical states (DOS) vanishes. 
To probe the DOS, light emitters are positioned inside the crystal.~\cite{Koenderink2002PRL,Lodahl2004Nature,Ogawa2004Science, Leistikow2011PRL} 
Such experimental realizations are difficult and sometimes lack appropriate sources as well as detection methods. 
On the other hand, a band gap can be indicated by a stop band in a directional experiment as shown by a peak in reflectivity or a trough in transmission.~\cite{Fleming1999OptLett,Blanco2000Nature,Noda2000Science,Vlasov2000Nature,Subramania2004ApplPhysLett,Schilling2005ApplPhysLett,Staude2010OptLett} 
Remarkably, a peak in reflectivity or a trough in transmission may also occur when an incident wave does not couple to a field mode inside the crystal.~\cite{Joannopoulos2008Book,Robertson1992PRL,Sakoda1995PRB} 
Thus, the experimentally observed stop bands are typically interpreted by a comparison with stop gaps from calculated band structures. 
Unfortunately, however, band structures pertain only to infinite and perfect crystals, hence features related to finite-size or to unavoidable deviations from perfect periodicity are not considered. 
A recent experimental study of silicon inverse woodpile photonic crystals observed the largest solid angle for which a broad photonic stop band has been reported.~\cite{Huisman2011PRB} 
It was asserted that intense reflection peaks measured with a large numerical aperture provide a faithful signature of the 3D photonic band gap. 
The limited reflectivity was attributed to the limited crystal thickness in comparison to the Bragg attenuation length and to surface roughness, although no theoretical or numerical support was offered for these notions.  

In the present article, we study numerically the reflectivity of 3D photonic band gap crystals to interpret their experimental realizations. 
We apply the finite element method for reflectivity calculations on crystals with the cubic diamond-like inverse woodpile structure. 
The high-index backbone of the crystals has a dielectric function similar to silicon. 
We investigate crystals with thicknesses up to ten unit cells.
Since the crystals are surrounded by vacuum, they have a finite support as in the experiments. 
We assess the previously invoked~\cite{Huisman2011PRB} limitations to the reflectivity, such as crystal thickness, angle of incidence, and Bragg attenuation length. 
Consequently, our numerical study provides an improved interpretation of reflectivity as a signature of a full 3D photonic band gap.

\section{Methods}
\begin{figure}[tbp]
\includegraphics[scale=0.30]{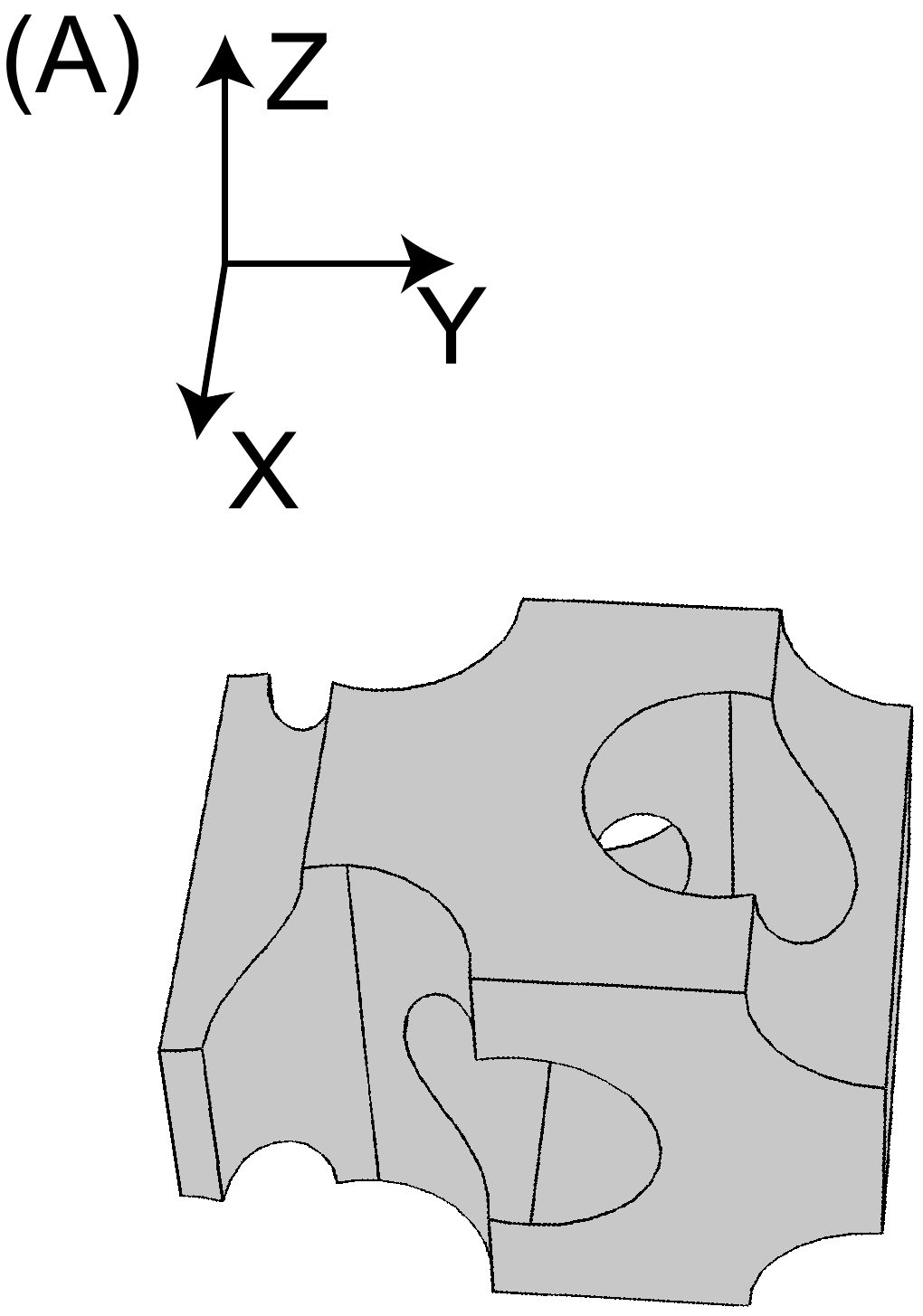}
\includegraphics[scale=0.25]{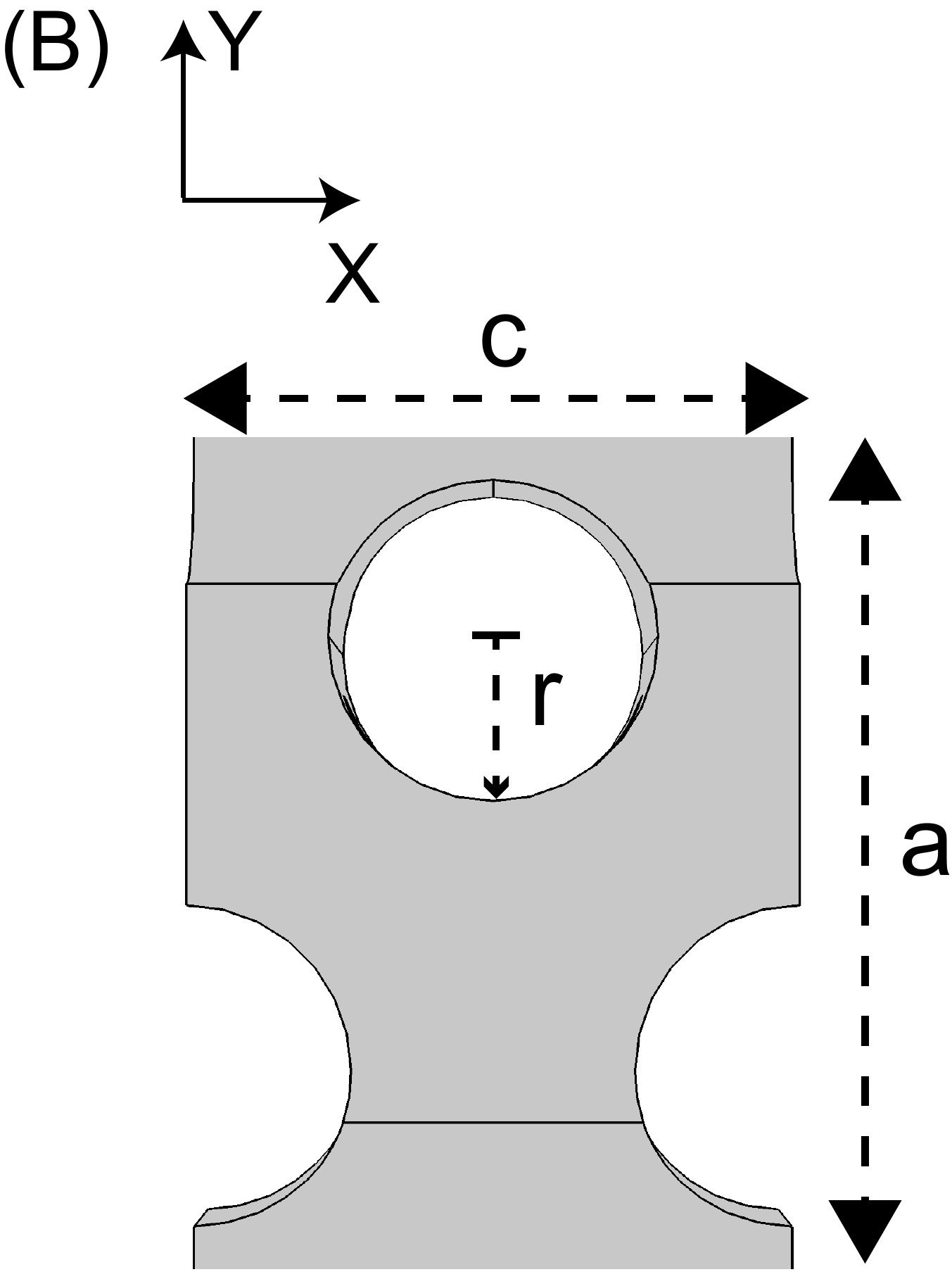}
\hspace{0.1\textwidth}
\includegraphics[scale=0.25]{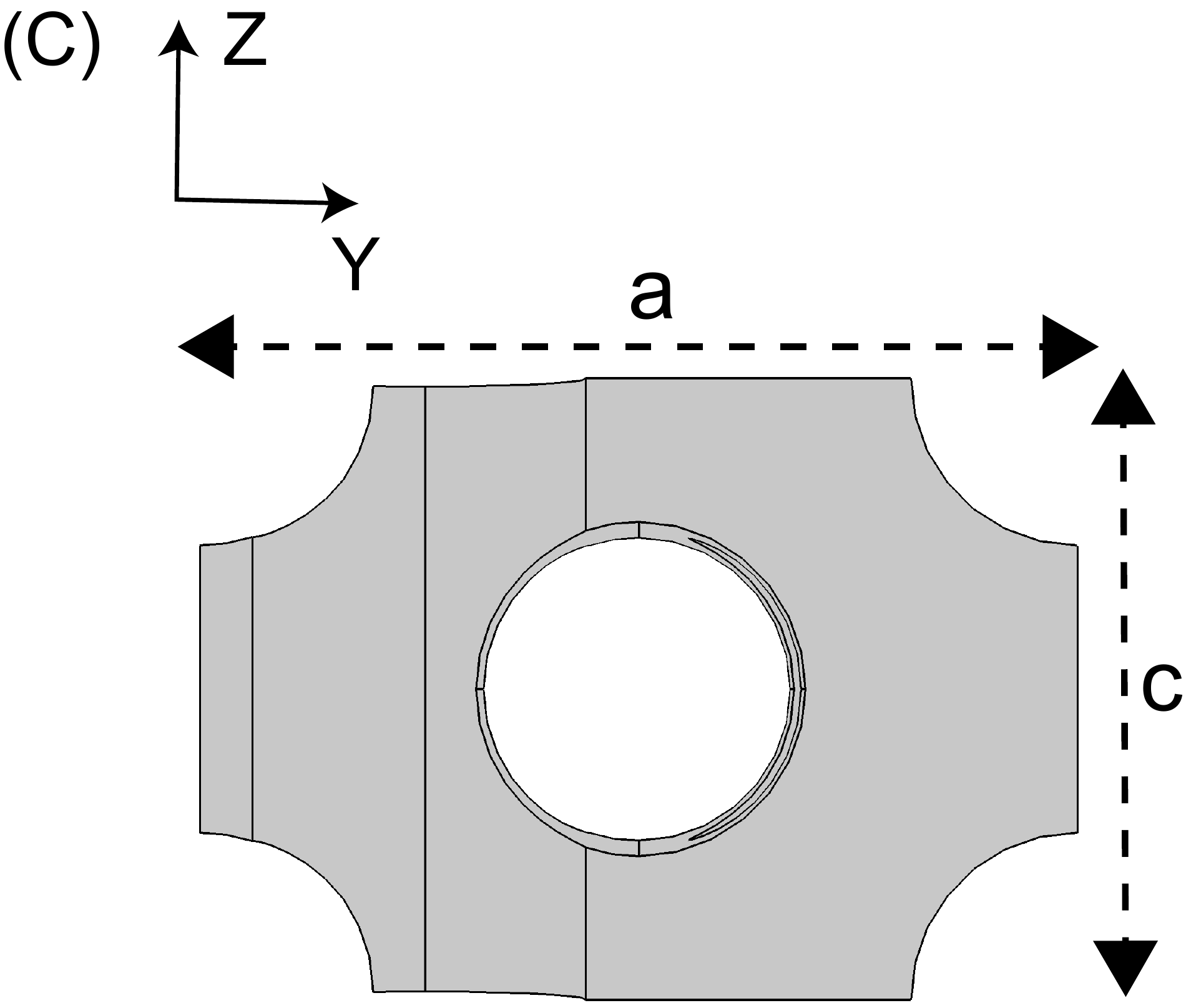}
\caption{The primitive unit cell of the cubic inverse woodpile photonic crystal structure. 
(A) Perspective view of the unit cell with the $XYZ$ coordinate system.The $X$-axis and the $Z$-axis are parallel to the two sets of pores. 
(B) View of the unit cell along the $Z$-axis with the lattice parameters $a$ and $c$ and pore radius $r$. 
(C) View of the unit cell along the $X$-axis.}
\label{fig:primitive-unit-cell}
\end{figure}

The primitive unit cell of the cubic inverse woodpile photonic structure is illustrated in Fig.~\ref{fig:primitive-unit-cell}.  
The crystal structure consists of two 2D arrays of identical pores with radius $r$ running in two orthogonal directions $X$ and $Z$. 
Each 2D array has a centered-rectangular lattice with lattice parameters $c$ and $a$. 
When the lattice parameters have a ratio $\frac{a}{c} = \sqrt{2}$, the diamond-like structure is cubic. 
Cubic inverse woodpile photonic crystals with $\epsilon = 12.1$ - typical of silicon - have a broad maximum band gap width $\Delta \omega/\omega_{c} = 25.4~\%$ relative to the central band gap frequency $\omega_{c}$ for pores with a relative radius $\frac{r}{a}=0.245$.~\cite{Woldering2009JAP,Hillebrand2003JAP} 

The $(X,Y,Z)$ coordinate system introduced in Fig.~\ref{fig:primitive-unit-cell}(A) is adopted in this paper. 
In terms of the conventional non-primitive cubic unit cell of the diamond structure, the $X$-axis has unit vector $a_{1} = \frac{1}{\sqrt{2}}[1~0~1]$, the $Y$-axis $a_{2} = [0~1~0]$, and the $Z$-axis $a_{3} = \frac{1}{\sqrt{2}}[\bar{1}~0~1]$.\cite{Ashcroft1976Book}
These crystals have been realized elsewhere in several different backbone materials using various techniques.~\cite{Santamaria2007AdvMater,Hermatschweiler2007AdvFunctMater,Jia2007JApplPhys,Tajiri2015APL} 
Our research group has fabricated 3D inverse woodpile photonic crystals from silicon using several CMOS-compatible methods.\cite{Woldering2008Nanotechnology, vandenBroek2012AFM,Grishina2015Nanotechnology} 

\begin{figure}[tbp]
\includegraphics[width=1.0\columnwidth]{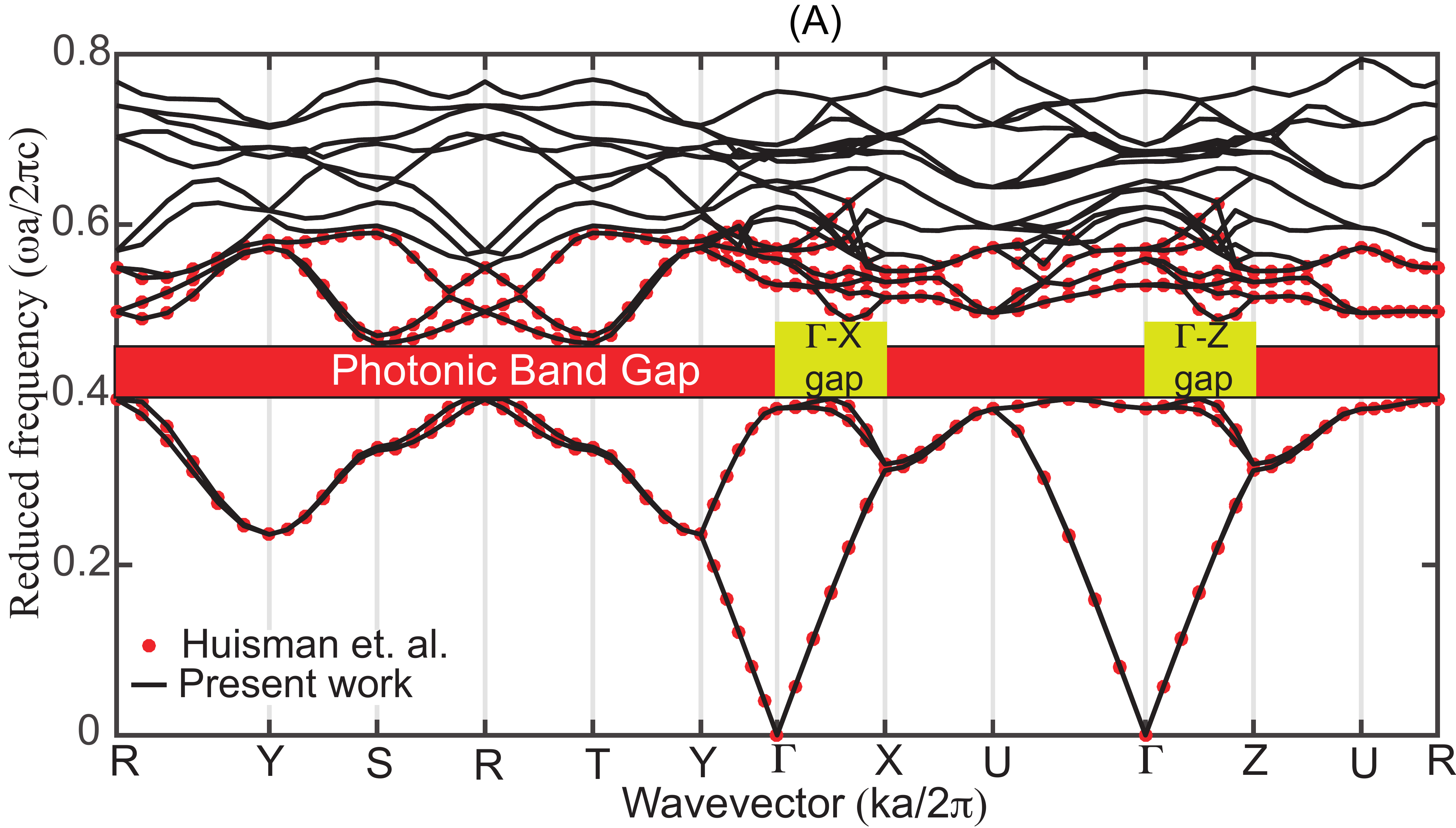}
\label{fig:Bandstructure}
%
\includegraphics[scale=0.25]{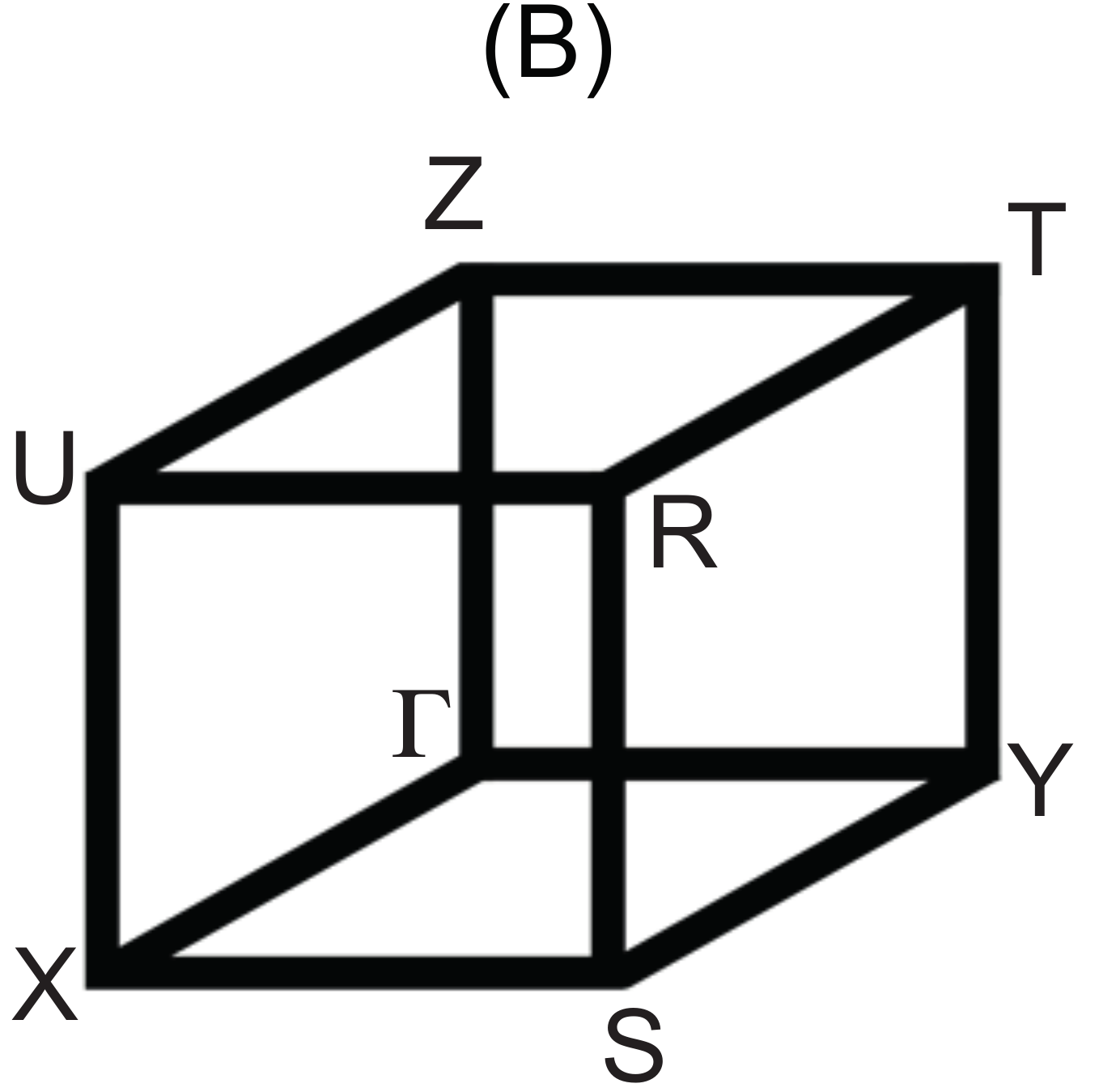}
%
\caption{
(A) Photonic band structure for the 3D inverse woodpile photonic crystal with $\frac{r}{a} = 0.19$ and $\epsilon_{Si}=12.1$. 
The red dots are from Ref.~\cite{Huisman2011PRB}. 
The black lines are our results calculated with a $2 \times 2 \times 2$ times greater 3D spatial resolution than in Ref.~\cite{Huisman2011PRB}, and with thirty bands instead of eight. 
The red bar marks the 3D photonic band gap, and the yellow bars mark stop gaps in the $\Gamma X$ and $\Gamma Z$ directions. 
Since the $\Gamma X$ stop gap is symmetry-related to the $\Gamma Z$ stop gap, we effectively consider both stop gaps in the present study.
(B) First Brillouin zone showing the high symmetry points and the origin at $\Gamma$. }
\label{fig:PhotonicBandstructure}
\end{figure}

In order to compute harmonic modes of Maxwell's equations in infinitely extended periodic dielectric structures for wave vectors in the first Brillouin zone, we employed the MPB plane-wave expansion method. ~\cite{Johnson2001OptExpress} 
Fig.~\ref{fig:PhotonicBandstructure}(A) and Fig.~\ref{fig:PhotonicBandstructure}(B) show the first Brillouin zone and the band structure respectively for an inverse-woodpile crystal with a non-optimal pore size. 
A broad photonic band gap (red bar) with a $15.2~\%$ relative width appears from the reduced frequency $\tilde{\omega}_{1}~=~0.395$ (bounded by $3^{rd}$ and $4^{th}$ bands) to $\tilde{\omega}_{2}~=~0.460$ ($5^{th}$ band). 
Throughout this paper, we express frequency as a reduced frequency $\tilde{\omega} = \omega a / (2 \pi c'))$, with $\omega$ the frequency, $a$ the lattice parameter, $c'$ the speed of light (not to be confused with the $c$ lattice parameter). 
Our definition $\tilde{\omega}$ for reduced frequency will be expressed in units of $(a/\lambda)$.  
The band structure shows two stop gaps in the $\Gamma Z$ direction, which is symmetry-related with the $\Gamma X$ direction in k-space. 
The lowest-frequency narrow stop gap appears between $\tilde{\omega} = 0.311$ and $\tilde{\omega} = 0.319$ and closes when moving in the $ZU$ direction. 
The second stop gap between $\tilde{\omega} = 0.395$ and  $\tilde{\omega} = 0.488$ is part of the 3D complete photonic band gap and it is much broader with $20.4~\%$ relative bandwidth. 
We find an excellent agreement between our calculations and earlier calculations in Ref.~\cite{Huisman2011PRB}. 
In the limit of $\omega \rightarrow 0$, we derive from the slope of the band in the band structure the estimated effective refractive index of the crystal equal to 2.28.

To accurately model the light propagation inside photonic band gap crystals with finite support, we employ a finite-element (FEM) based commercial solver~\cite{COMSOLMultiphysics} for the time-harmonic Maxwell equations. 
This solver has the ability to efficiently sample complex geometrical features often encountered in photonic crystals. 

\begin{figure}[tbp]
\includegraphics[width=1.0\columnwidth]{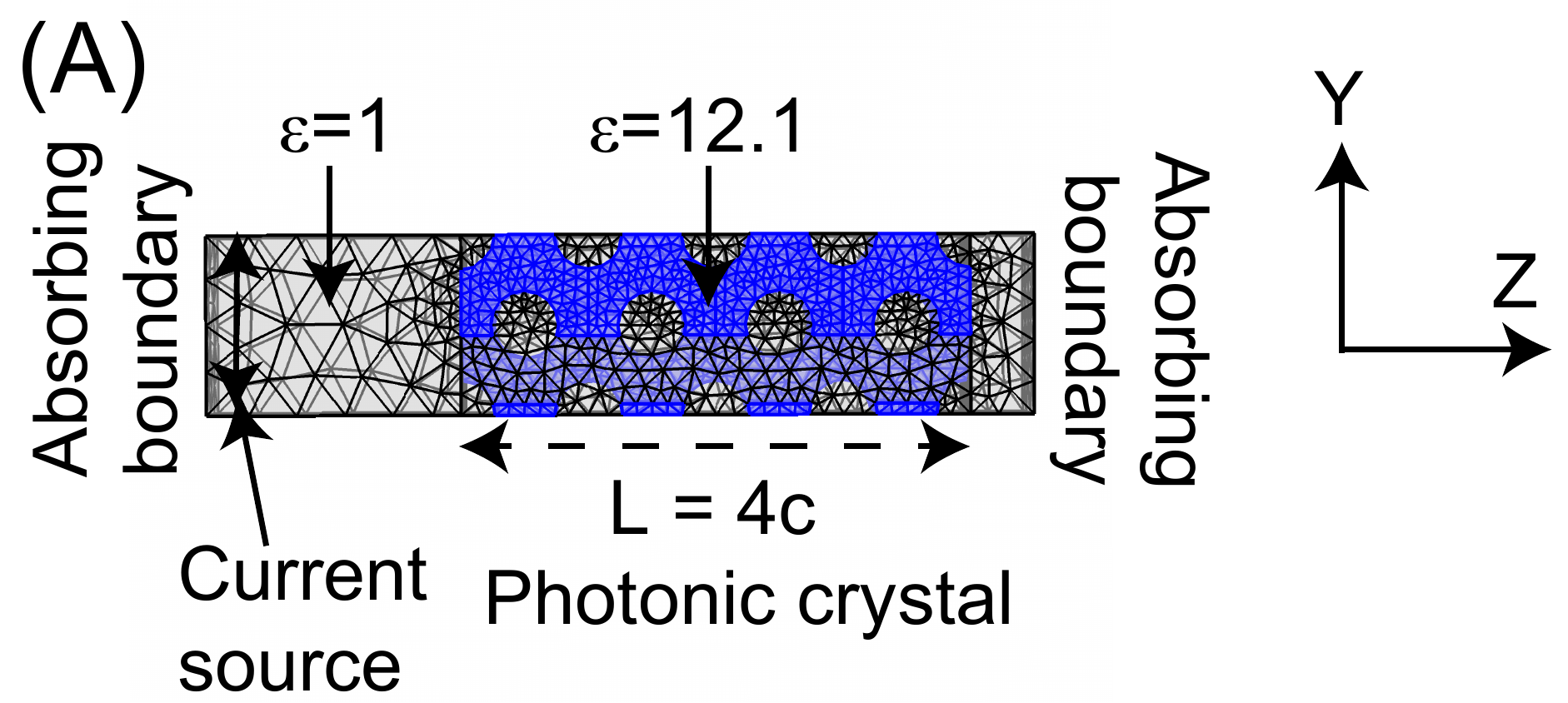}
%
\hspace{0.1\textwidth}
\includegraphics[width=1.0\columnwidth]{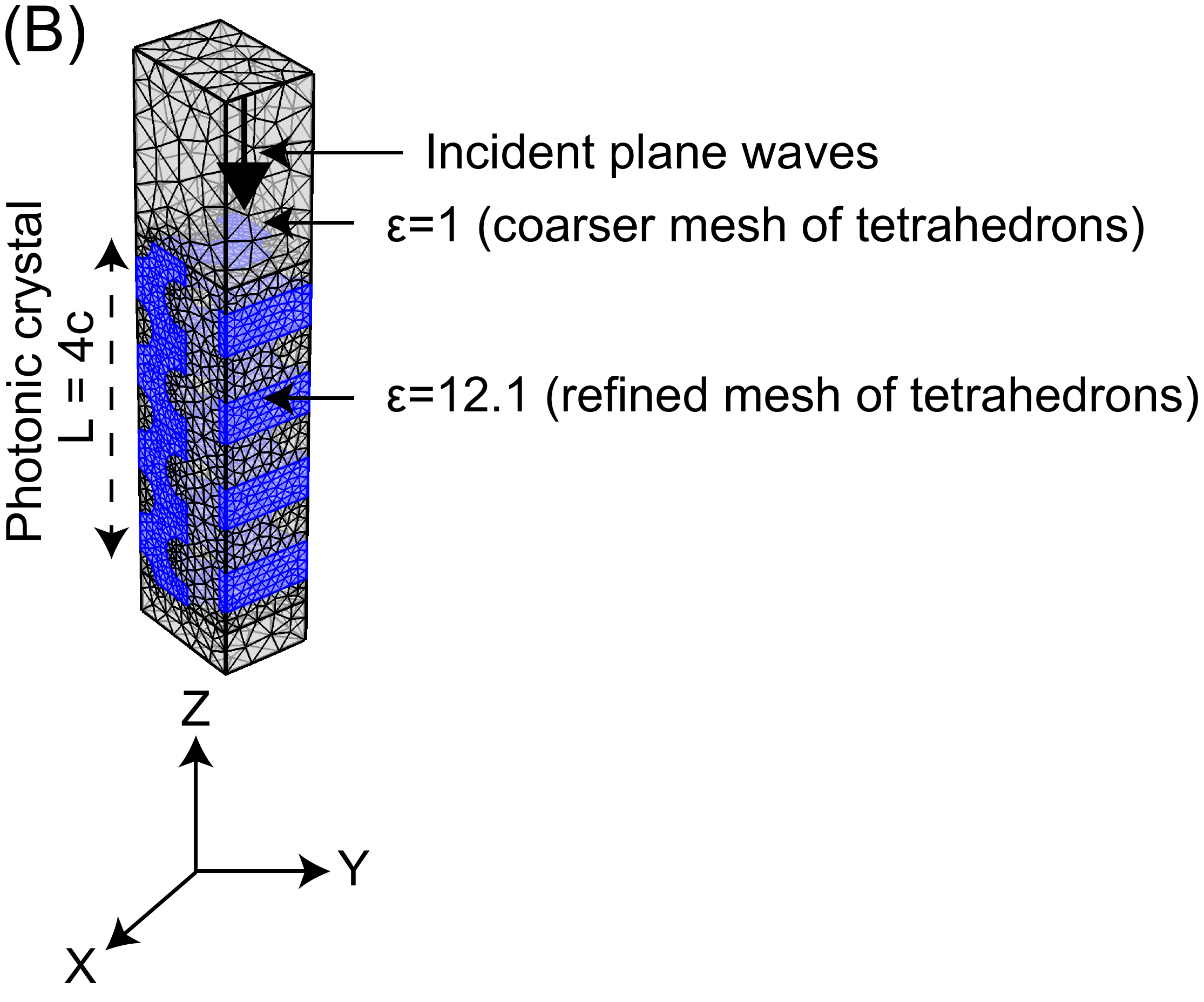}
%
\caption{Illustration of the computational cell for a photonic crystal with thickness $L$ = $4c$. 
(A) View along the X-direction $[101]$. 
The source of plane waves is at the left, and is separated by an air layer from the crystal. 
The computational cell is bounded by absorbing boundaries at -Z and +Z, and by periodic boundary conditions on $\pm$X and $\pm$Y 
(B) Perspective view of the computational cell.}
\label{fig:ComputationalCell}
\end{figure}
Our numerical scheme for a crystal with finite support is employed to calculate the reflectivity and the transmission spectra. 
For comparison of our calculations with experimental results, we choose the lattice parameter to be $a = 690$ nm.~\cite{vandenBroek2012AFM} 
Following Refs.~\cite{Woldering2009JAP,Huisman2011PRB}, a dielectric permittivity $\epsilon = 12.1$ is adopted, typical for silicon in the near infrared and telecom ranges. 
Figure~\ref{fig:ComputationalCell}(A) illustrates the view of the computational cell along the X-direction. 
The current source, at the left, is separated from the crystal by an air layer. 
The current source emits plane waves with either s-polarization (electric field normal to the plane of incidence) or p-polarization (magnetic field normal to the plane of incidence), and at different angles of incidence. 
To mimic infinite space, absorbing boundaries are employed in the directions where the crystal is finite in size. 
Absorbing boundaries at $-Z$ and $+Z$ in Fig.~\ref{fig:ComputationalCell}(A) minimize the back reflections. 
We employ Bloch-Floquet periodic boundaries in the $\pm X$ and the $\pm Y$ directions to describe a crystal slab.~\cite{Joannopoulos2008Book} 

Figure~\ref{fig:ComputationalCell}(B) illustrates the finite element mesh used to subdivide the 3D computational cell. 
Since a tetrahedron can mesh any 3D volume regardless of shape or topology~\cite{Jin1993Book}, we used tetrahedron as the basic element in our finite element mesh. 
A limit of $\frac{\lambda_{0}}{8\sqrt{\epsilon}}$ is imposed to the edge length on any tetrahedron, where $\lambda_{0}$ is the smallest wavelength of the incident plane wave in vacuum and $\epsilon$ is the dielectric permittivity of the medium. 
The finite element mesh consists of 27852 tetrahedra per crystal unit cell.  
Since the edge length of the tetrahedon used to mesh the geometry is sufficiently smaller than the wavelength, the tetrahedron polynomial basis functions resolve the waves resulting in converging solution.~\cite{Ihlenburg1995ComputersMathApplic} 
A refined mesh is used at the interface between the high-index material and the low-index material to reduce dispersion errors. 
In order to achieve computational efficiency in solving the time-harmonic Maxwell's equations, we apply a direct solver named MUMPS. 
This solver is fast, multi-core capable and cluster capable. 
For a single frequency and a single angle of incidence, the computational time is 35 s on a Intel Core i7 machine with a single processor of 4 cores. 
We found that the computational time increases sub-linearly with respect to the number of frequency steps and the number of angle of incidence steps. 

\begin{figure}[tbp]
\centering
\includegraphics[width=1\columnwidth]{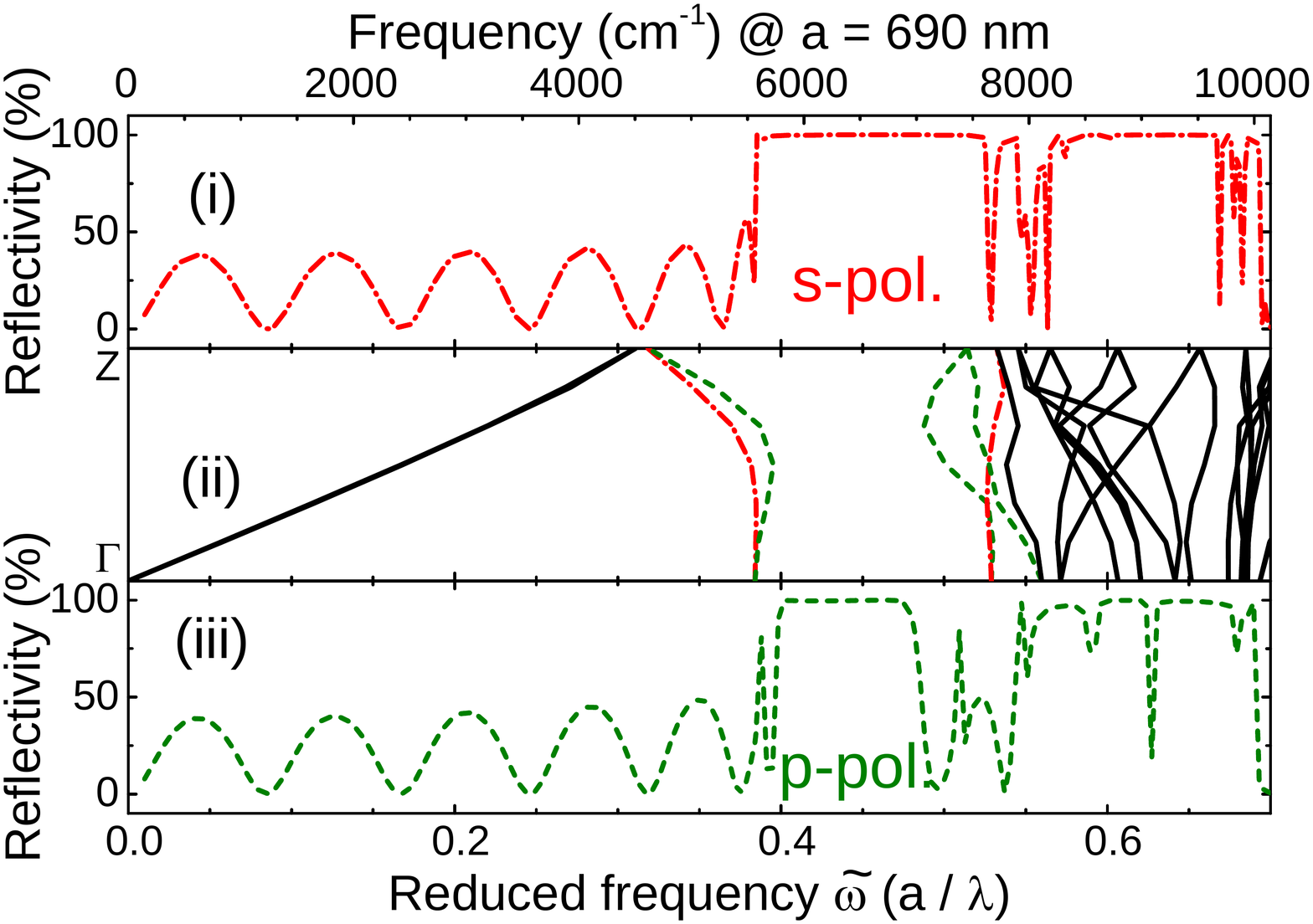}
\caption{Calculated reflectivity spectra for an inverse woodpile photonic crystal along the $\Gamma Z$ high symmetry direction in k-space. 
The red curve in panel (i) and the green curve in panel (iii) are the reflectivity spectra for s- and p-polarization, respectively. 
The corresponding band structure for the $\Gamma Z$ direction is shown in panel (ii). 
The frequency ranges of the s- and p-stop bands agree excellently with corresponding stop gaps in the photonic bandstructure. 
The assignment of the bands is based on Figure~\ref{fig:reflectivity-and-bandstructure_zoomed-in}.} 
\label{fig:reflectivity-and-bandstructure}
\end{figure}

\begin{figure}[tbp]
\centering
\includegraphics[width=1.0\columnwidth]{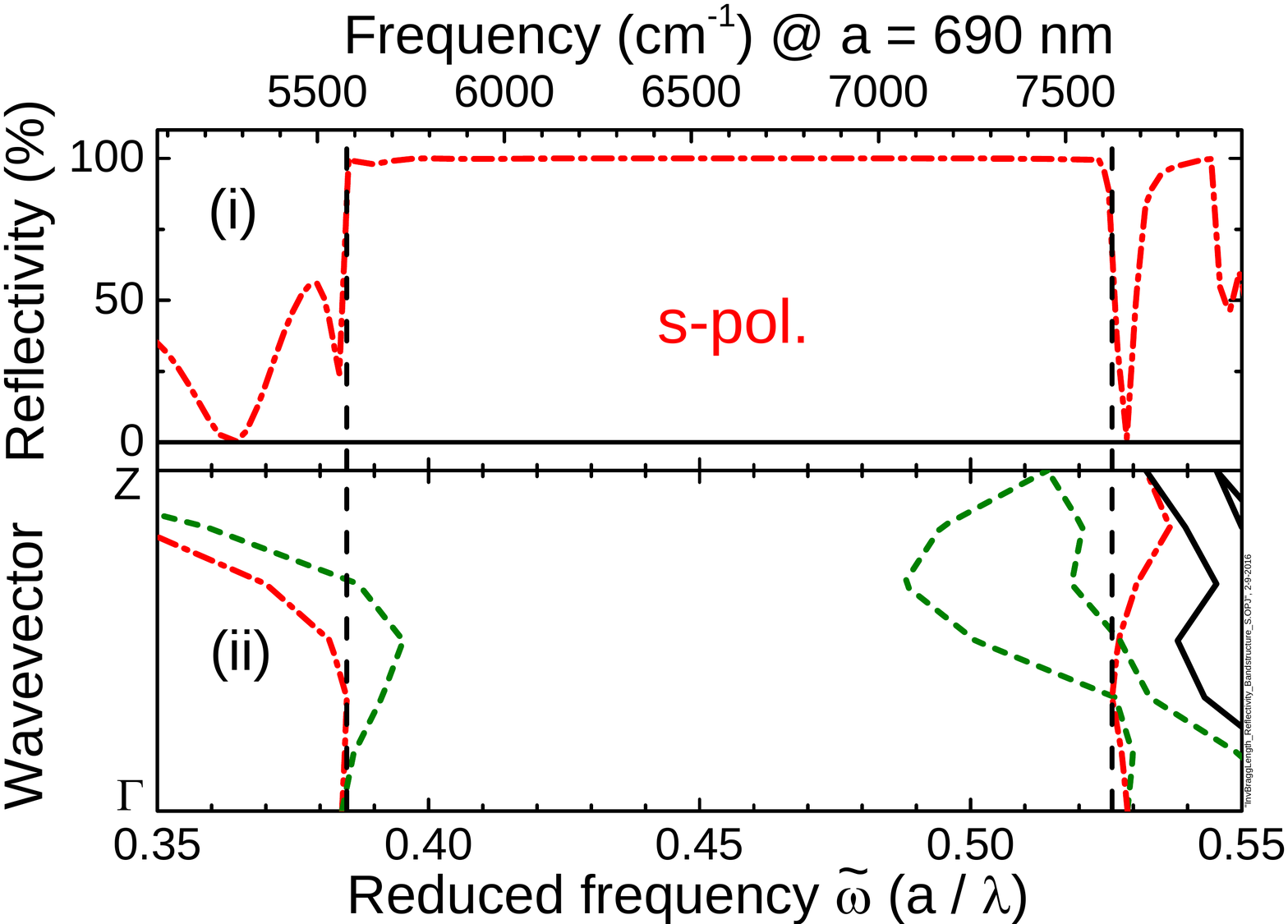}

\includegraphics[width=1.0\columnwidth]{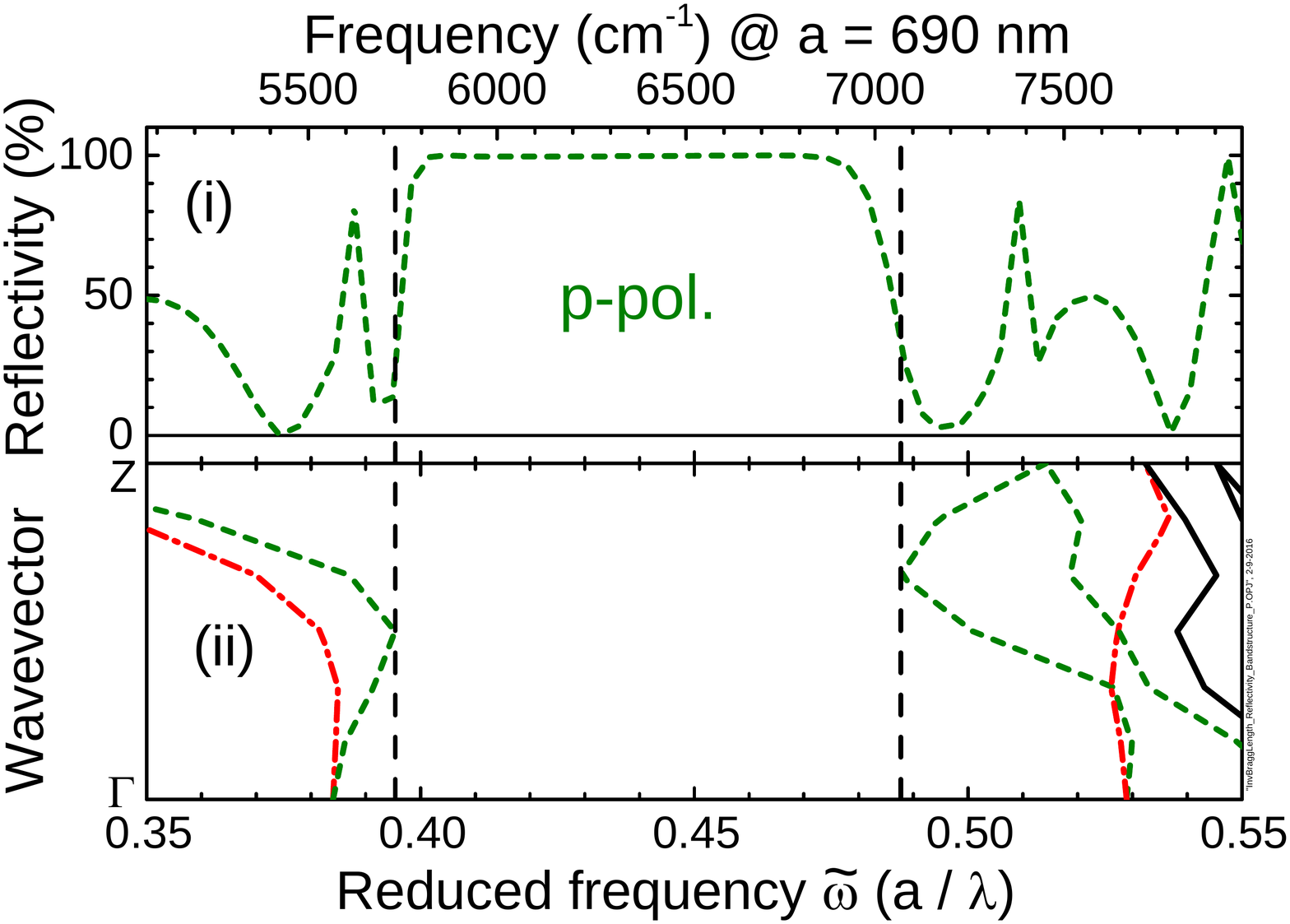}

\caption{Zoom-in near the stop band of (i) the calculated reflectivity spectrum for s- and p-polarization and (ii) the wavevectors of the photonic bands along the $\Gamma Z$ high symmetry direction in k-space. 
The vertical black dashed lines indicate the edges of the stop band (i) and the matching stop gap edges (ii). 
The top ordinate gives the frequency in wavenumbers ($cm^{-1}$) for a lattice parameter $a = 690$ nm as in the experiments.\cite{Huisman2011PRB} 
Near the stop gap edge, we identify s-bands (red dashed-dotted curves) and p-bands (green dashed curves). } 
\label{fig:reflectivity-and-bandstructure_zoomed-in}
\end{figure}

\section{Results}

\subsection{Frequency-resolved reflectivity}
We performed an extensive set of reflectivity calculations on crystals with finite support. 
Figure~\ref{fig:reflectivity-and-bandstructure} shows representative spectra for a thin crystal with a thickness $L = 4c$.
We consider polarization-resolved (either s or p) plane waves that are incident normal to the crystal surface and travel along the $\Gamma Z$ high symmetry direction in $k$-space.
The polarization- and frequency- resolved spectra in Fig.~\ref{fig:reflectivity-and-bandstructure} ((i), (iii)) reveal Fabry-P{\'e}rot fringes that correspond to standing waves in the finite crystal slab.  
The strong reflectivity peaks near $\tilde{\omega} = 0.45$ indicate stop bands for both s- and p-polarizations. 
The p-stop band appears between $\tilde{\omega} = 0.395$ and $\tilde{\omega} = 0.488$, with a broad relative bandwidth $20~\%$. 
The s-stop band appears between $\tilde{\omega} = 0.385$ and $\tilde{\omega} = 0.526$ and it is about $1.5 \times$ broader (relative bandwidth $31~\%$) than the p-stop band. 
At frequencies beyond $\tilde{\omega}=0.55$ several bands of high reflectivity appear for which we have currently no interpretation; in these frequency bands the band structures reveal extremely complex couplings of multiple Bragg conditions~\cite{Vos2000PhyLettA2000} that lead to complex band structures that are sometimes also referred to as "\emph{spaghetti-like}" behavior.

The frequency ranges of the s- and the p-stop bands agree very well with corresponding stop gaps in the photonic band structure. 
As a result, we can assign bands in the band structure near the stop bands to have dominantly s- or p-character, as shown in the zoomed-in Figure~\ref{fig:reflectivity-and-bandstructure_zoomed-in}. 
Since the $3^{rd}$ photonic band at the lower stop gap edge (near $\tilde{\omega} = 0.385$) agrees with the lower boundary of the s-stop band, we conclude that this band has dominantly s-character. 
Furthermore, the $4^{th}$ band is located inside the s-stop band and agrees with the lower edge of the p-stop band at $\tilde{\omega} = 0.395$.
Therefore, we conclude that this band must have dominantly p-character. 
Near the upper gap edge, the $7^{th}$ band near $\tilde{\omega} = 0.526$ agrees with the upper s-stop band edge and is thus likely an s-band. 
The $5^{th}$ and $6^{th}$ bands between $\tilde{\omega} = 0.49$ and $\tilde{\omega} = 0.526$ are situated well inside the s-stop band and can therefore only have p-character; indeed, these bands lie outside the p-stop band. 
This assignment of bands 5, 6, and 7 is further supported by the observation that band 7 crosses bands 5 and 6 at $\tilde{\omega} = 0.526$, without revealing avoided crossings. 

Figures~\ref{fig:reflectivity-and-bandstructure} and~\ref{fig:reflectivity-and-bandstructure_zoomed-in} also reveal that even a thin crystal with a thickness of only four unit cells has a maximum reflectivity of $99.99~\%$, very close to ideal $100~\%$. 
The high reflectivity implies that inverse woodpile crystals interact strongly with light. 
The strong reflectivity peaks also imply that the maximum reflectivity observed in Ref.~\cite{Huisman2011PRB} is most likely not limited by finite size, since there the crystals were even thicker ($L = 12 c$). 
At this time, we surmise that the measured reflectivity was limited by roughness of the crystal-air interface, and by roughness inside the pores. 

\begin{figure}[tbp]
\includegraphics[width=1\columnwidth]{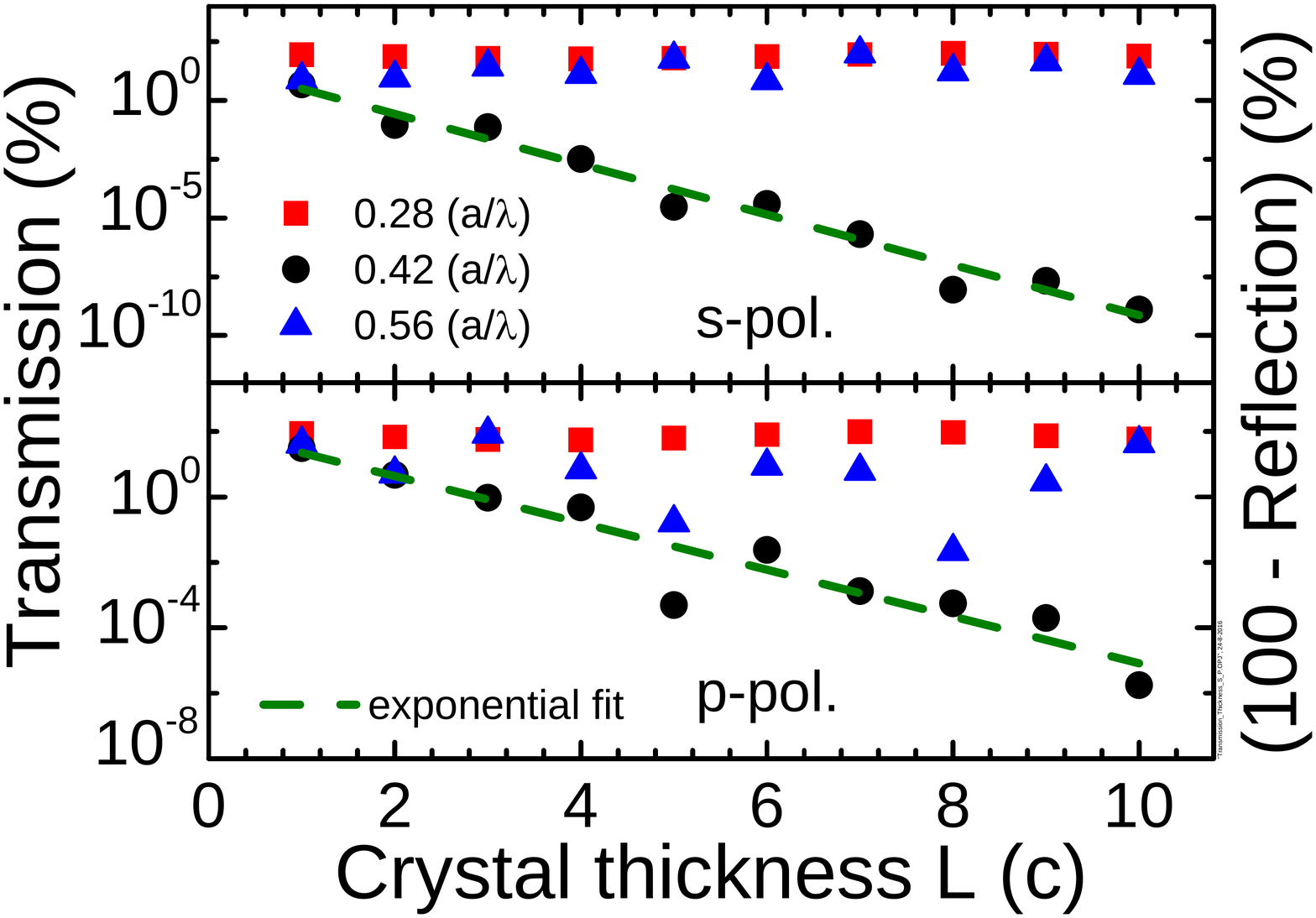}
\caption{
Transmission versus thickness for a silicon inverse woodpile photonic crystal in the $\Gamma Z$ direction for s- (top) and p-polarizations (bottom). 
Red squares, black circles and blue triangles pertain to frequencies below, inside, and above the stop gap, respectively. 
The green dashed lines are the exponential decay of transmission with crystal thickness at frequencies in the stop gap (Eq.~(\ref{eq:transmission_BraggLength})). 
}
\label{fig:transmission-vs-thickness}
\end{figure}

\subsection{Finite-size effects: Bragg attenuation length}
To investigate the effect of finite thickness of the crystal, we calculated the transmission for thicknesses in the range from $L=1c$ to $10c$.
Figure~\ref{fig:transmission-vs-thickness} shows that for a given frequency inside the stop gap, the transmission decays exponentially for both s- and p-polarizations. 
For frequencies below or above the stop gap, the transmission is nearly constant, with some small variations with crystal thickness that are the result of the Fabry-P{\'e}rot fringes that vary with crystal thickness, as is well-known for 1D Bragg stacks.~\cite{Yariv1984Book,Yeganegi2014PRB}

Inside a stop gap the complex wave vector $k$ has a nonzero imaginary component $Im(k)$ since the waves are damped by Bragg diffraction interference.~\cite{Yariv1984Book} Thus, at frequencies in a stop band we write the transmission $T$ as
\begin{equation} 
T(\omega) = exp\left({-\frac{L}{L_{B}(\omega)}}\right),
\label{eq:transmission_BraggLength}
\end{equation} 
with $L_{B}$ the Bragg attenuation length equal to 
\begin{equation} 
L_{B}(\omega) = \frac{1}{Im(k)(\omega)}.
\label{eq:BraggLength}
\end{equation} 
The Bragg attenuation length gives the distance covered by incident light until it has exponentially decayed to a fraction $1/e$ as a result of Bragg interference. 
%
Figure~\ref{fig:transmission-vs-thickness} reveals that even inside the stop band the transmission shows modulations, as was previously identified in $1D$ stacks.~\cite{Yariv1984Book, Yeganegi2014PRB} 
The reason is that transmission also contains the effects of both front and back crystal surfaces. 
Nevertheless, the envelope of the transmission is exponential as in Eq.~(\ref{eq:transmission_BraggLength}). 
 
The Bragg attenuation length is usually expressed in terms of the distance between lattice planes $d_{hkl}$. 
Therefore, we reduce the Bragg length to the $\left\lbrace hkl = 220 \right\rbrace$ lattice spacing $d_{220}$ that is equal to $d_{220}~=~c/2$. 
The s-polarized data in Fig.~\ref{fig:transmission-vs-thickness} agree well with an exponential fit (Eq.~(\ref{eq:transmission_BraggLength})) with a slope that yields a Bragg attenuation length $L_{B} = 0.74 d_{220}$. 
For the p-polarized data in Fig.~\ref{fig:transmission-vs-thickness}, we obtain a Bragg attenuation length $L_{B} = 1.21 d_{220}$ at the gap center, which is about $1.5 \times$ larger than for s-polarization at the gap center. 
This observation agrees quantitatively with the reflectivity spectrum where the s-polarized stop band is also $1.5 \times$ broader than the p-polarized stop band (see Fig.~\ref{fig:reflectivity-and-bandstructure}). 

This behavior can be understood as follows: The Bragg attenuation length at the center frequency of a stop gap of a Bragg stack satisfies~\cite{Vos2015Chapter}
\begin{equation}
L_{B} = \frac{2 d}{\pi S} = \frac{2 d}{\pi} \frac{\omega_{c}}{\Delta \omega},
\label{eq:Bragg_length}
\end{equation}
where the photonic interaction strength $S$ is defined as the polarizability per volume of a unit cell~\cite{Vos1996PRB} that is estimated from the relative frequency band width of the stop band $S \approx \frac{\Delta \omega}{\omega_{c}}$.~\cite{Vos2015Chapter}
We find that the Bragg lengths exceed the earlier experimental estimate in Ref.~\cite{Huisman2011PRB} that was derived from the width of the stop bands by a factor $6$ to $9$. 
Hence, crystals with a thickness of 12 unit cells studied in these experiments are effectively in the thick crystal limit since $\frac{L}{L_{B}}=5$ to $8$.

\begin{figure}[tbp]
\includegraphics[width=1.0\columnwidth]{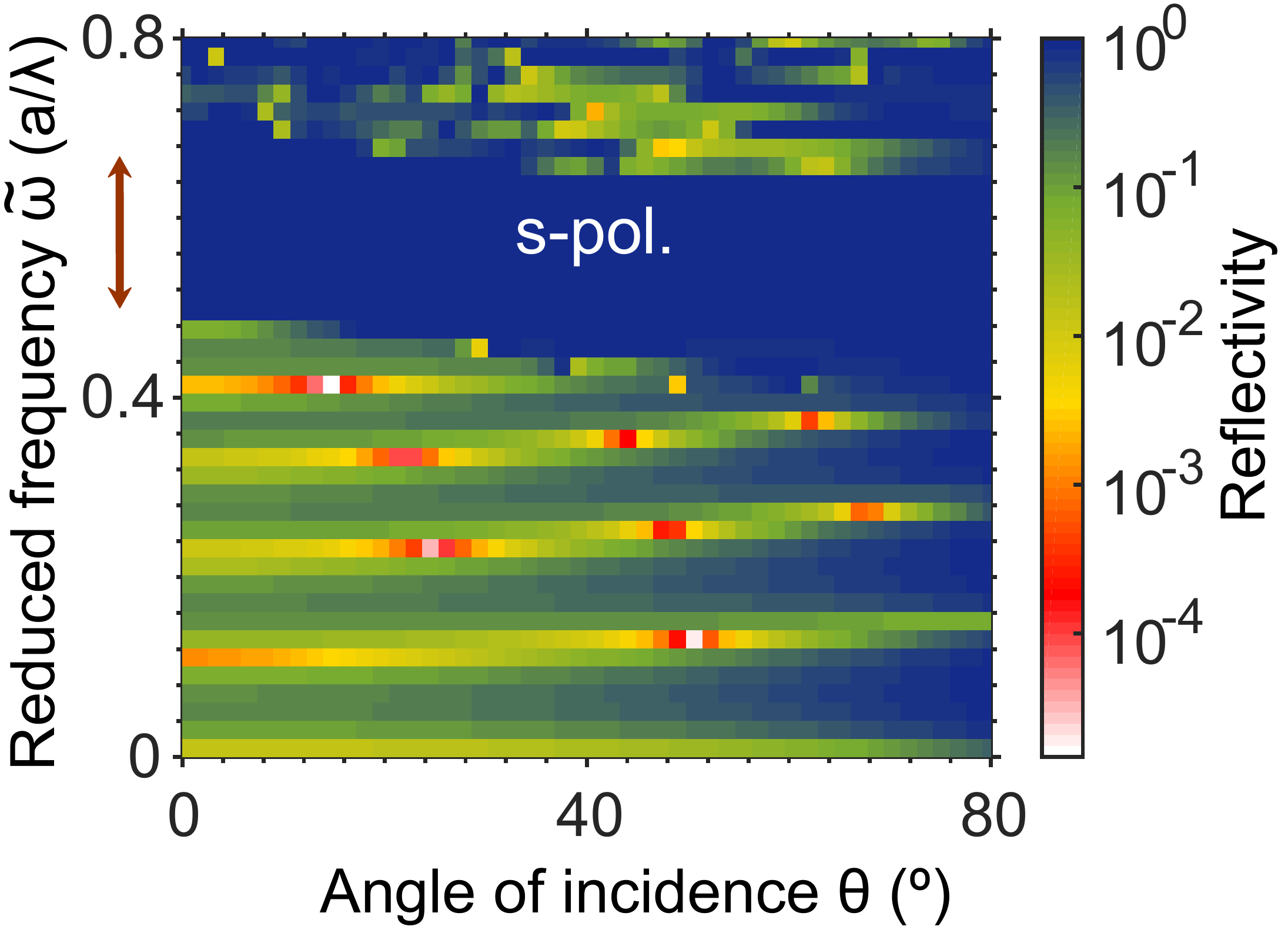}
\vspace{0.1\textwidth}
\includegraphics[width=1.0\columnwidth]{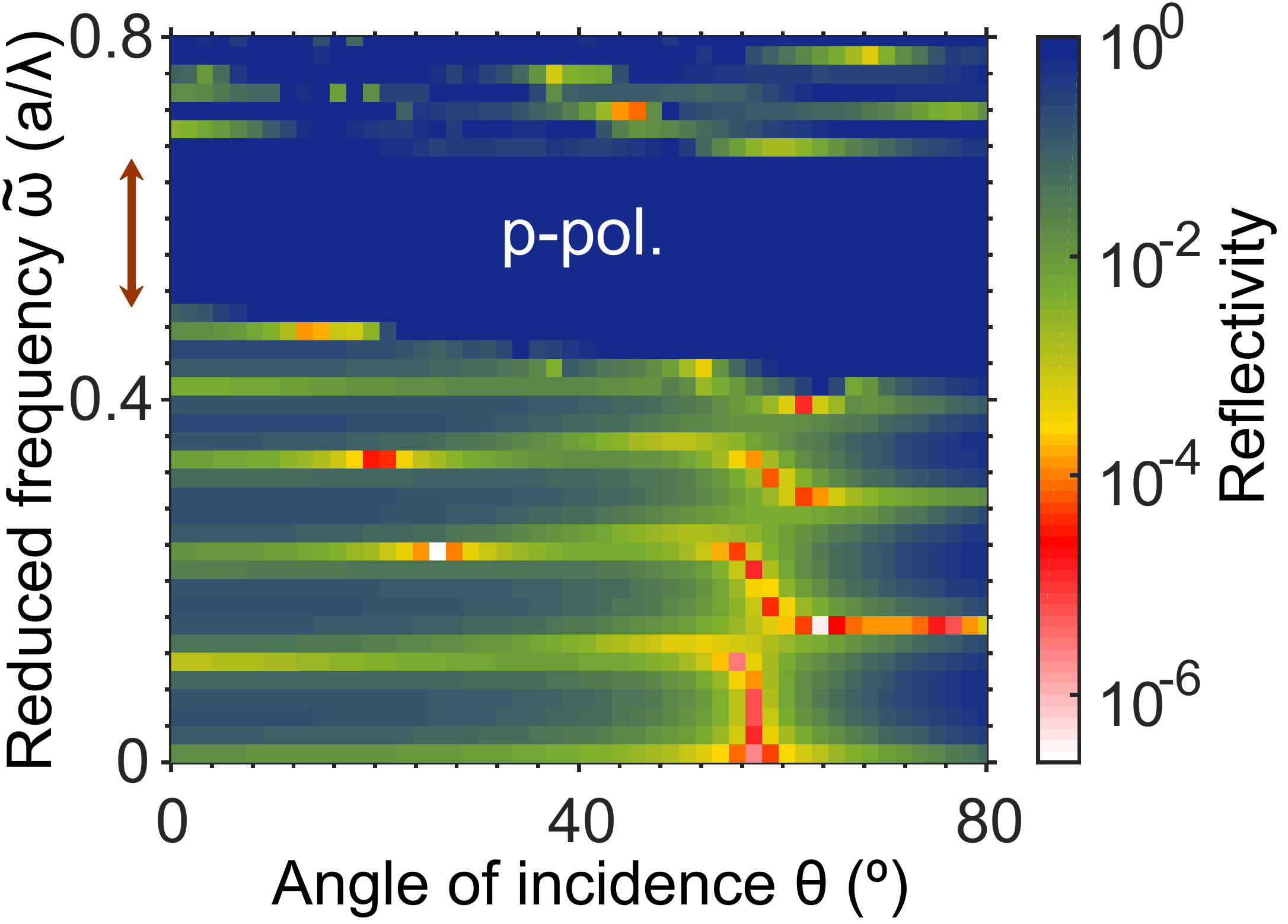}
\caption{Calculated angle- and frequency- resolved reflectivity spectra in the $\Gamma Z$ direction for a crystal with thickness $L=4c$ for (a) s-polarization and (b) p-polarization. 
The dark blue color represents high reflectivity of nearly $100~\%$ that occurs in the stop band at all angles. 
The white color represents minimal reflectivity near $0\%$, that occurs in the Fabry-P{\'e}rot fringes, at the Brewster angle, and in their hybrization in the range $56 \degree \leq \theta \leq 60 \degree$.
The brown double arrow represent the calculated stop gap in the $\Gamma Z$ direction (from Fig.~\ref{fig:PhotonicBandstructure}). }
\label{fig:AngleFrequencyReflectivity}
\end{figure}

\subsection{Angle- and frequency-resolved reflectivity}
As an extension of the reflectivity at normal incidence parallel to the $\Gamma Z$ direction, we have also calculated reflectivity spectra for angles of incidences up to $80 \degree$ off normal. 
Above, we performed calculations for non-optimal pore radius ($\frac{r}{a} = 0.19$) in order to interpret the experiments in Ref.~\cite{Huisman2011PRB}. 
Now we calculate reflectivity spectra for an inverse woodpile crystal with a maximum band gap width corresponding to an optimal pore size $\frac{r}{a}=0.245$. 
Following Ref.~\cite{Li1993JPCRD}, a midinfrared dielectric permittivity of silicon $\epsilon_{si}=11.68$ is adopted.
Figure~\ref{fig:AngleFrequencyReflectivity} shows the angle-resolved and frequency-resolved reflectivity spectra for an inverse-woodpile crystal with a thickness $L=4c$. 
Near $\tilde{\omega} = 0.6$ we observe broad stop bands with nearly $100\%$ reflectivity for both polarizations. 
The stop bands at normal incidence appear at a higher frequency than in Fig.~\ref{fig:reflectivity-and-bandstructure} since the air fraction is greater and hence the average index of the crystal is lower. 
The stop bands agree very well with the stop gaps calculated using the plane-wave expansion method (not shown). 
We observe that the frequency range of the stop bands hardly changes with angle of incidence, which is plausible since the stop bands are part of the 3D band gap. 
This result also supports the previous experimental notion~\cite{Huisman2011PRB} that intense reflectivity peaks collected with an objective with a large numerical aperture give a \emph{bona fide} signature of the 3D band gap. 
Fabry-P{\'e}rot fringes are visible for both polarizations corresponding to standing waves in the finite crystal. 

\begin{figure}
\includegraphics[width=1.0\columnwidth]{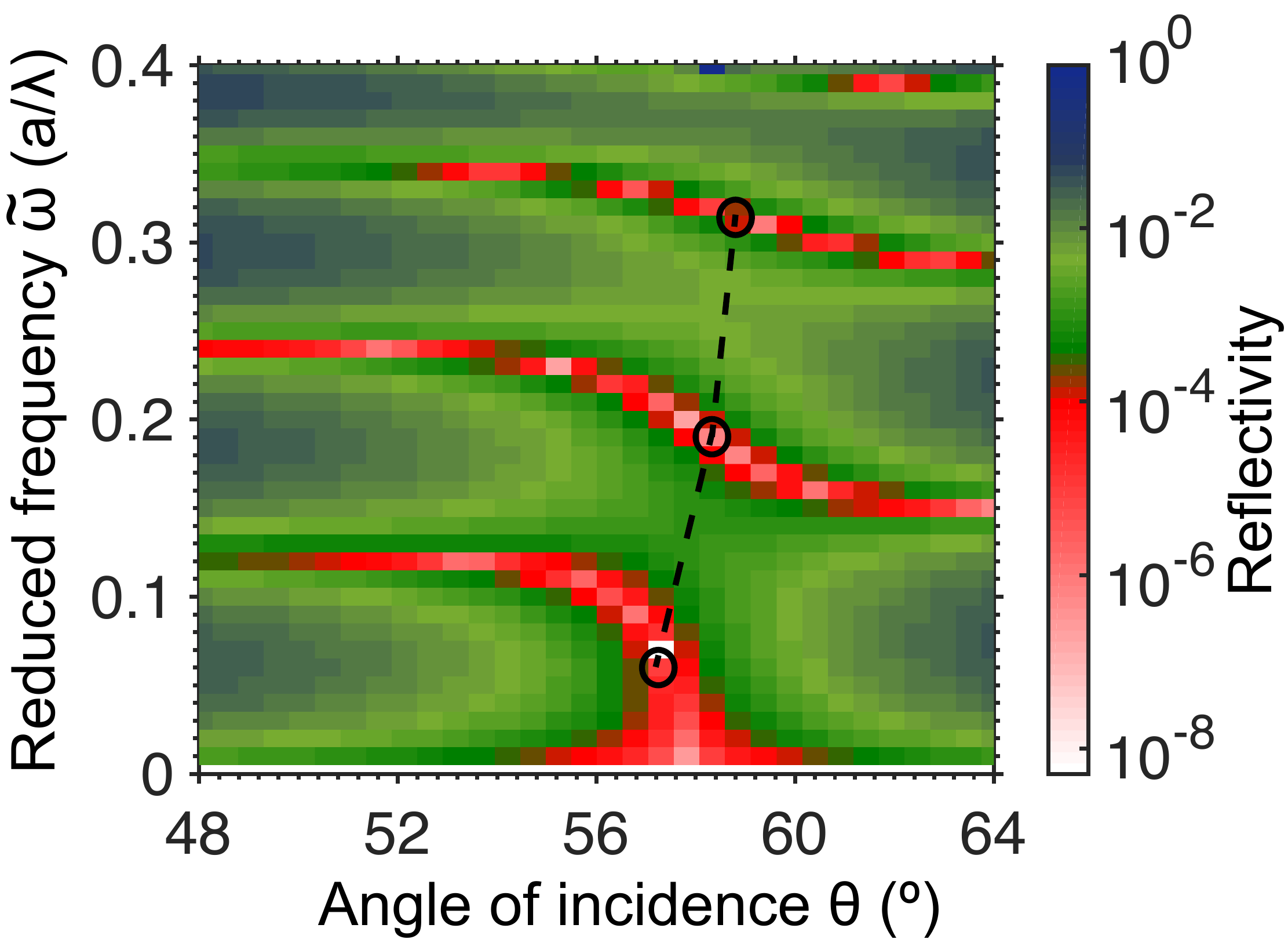}
\centering
\caption{Hybridization of the zero reflection of Fabry-P{\'e}rot fringes and the Brewster angle, shown by calculated angle- and frequency- resolved reflectivity spectra in $\Gamma Z$ direction for p-polarization. The black line is a guide to the eye that connects the halfway-points (circles) of the bends in the fringes.}
\label{fig:BrewsterEffect}
\end{figure}

For p-polarization, Fig.~\ref{fig:AngleFrequencyReflectivity} reveals an intriguing hybridization of the zero reflection of Fabry-P{\'e}rot fringes and the Brewster angle, which has not yet been observed in experiments. 
In order to characterize this feature, we calculated reflectivity spectra for p-polarized incident waves using a higher resolution in frequency and angle of incidence, shown in Fig.~\ref{fig:BrewsterEffect}. 
We note that the Fabry-P{\'e}rot fringes have a constant frequency for angles of incidence up to $\theta = 54\degree$ before bending. 
Beyond $\theta = 61\degree$, the fringes shift down in frequency to nearly the frequency of the lower order one at $\theta \leq 54\degree$, \emph{e.g.}, the $n = 2$ fringe at $\tilde{\omega}=0.24$ ($\theta \leq 54 \degree$) shifts to $\tilde{\omega}=0.15$ ($\theta > 61 \degree$), which is close to the frequency of the $n = 1$ fringe at $\theta \leq 54 \degree$. 
In the limit of $\omega \rightarrow 0$, we derive from the slope of the band in the band structure the estimated effective refractive index of the crystal equal to 1.68 and hence the Brewster angle $\theta_{B}=59.2\degree$. 
Since the angular range of bending occurs in the range ($56 \degree \leq \theta \leq 60 \degree$) of the Brewster angle ($\theta_{B}=59.2\degree$), we conclude that this bending is apparently a hybridization between the zero reflection of Fabry-P{\'e}rot fringes and the Brewster angle. 
The radius of curvature for a bend increases while approaching the stop band. 
A possible cause for the increase in radius of curvature with frequency may be the approach of the 3D photonic band gap, that prevents light from entering at a Brewster angle. \\

For comparison with a simple optical system, we have analytically computed the angle- and frequency-resolved reflectivity spectra of a thin film for p-polarization (see Appendix B). 
We find that the Brewster angle for a thin dielectric film is constant with frequency. 
We also observe that the Fabry-P{\'e}rot fringes have a constant frequency at all angles and do not bend near the Brewster angle. 
These observations on a thin film also pertain to a 1D Bragg stack as shown in Ref.~\cite{Shen2014Science}. 
Therefore, the hybridization between the zero reflection of Fabry-P{\'e}rot fringes and the Brewster angle appears to be a characteristic property of the 3D photonic crystal. 

\section{Discussion}

\subsection{Fabry-P{\'e}rot fringes and diffraction condition}

The polarization-resolved reflectivity spectra for the cubic diamond-like inverse woodpile structure in Fig.~\ref{fig:reflectivity-and-bandstructure} reveal Fabry-P{\'e}rot fringes that correspond to standing waves in the periodically layered finite crystal.
There are three corollaries based on theory for a periodic layered Bragg reflector~\cite{Yariv1984Book} with a thickness of $N$ unit cells.
First, a reflectivity peak occurs at the centers of the stop gaps. 
Second, between any two stop gaps there are exactly $(N - 1)$ troughs in the reflectivity spectra.  
Third, there are exactly $(N - 2)$ side lobes to the reflectivity peak. 

The bandstructure in Fig.~\ref{fig:reflectivity-and-bandstructure} (ii) shows two stop gaps in the $\Gamma Z$ direction. 
A narrow stop gap appears near $\tilde{\omega} = 0.311$ and the broad stop gap appears near $\tilde{\omega} = 0.39$. 
We now interpret the spectra in Fig.~\ref{fig:reflectivity-and-bandstructure} (i), (iii) for $N=4$ unit cells in terms of the 3 corollaries above. 
For s-polarization reflectivity in Fig.~\ref{fig:reflectivity-and-bandstructure} (i), we observe a peak near $\tilde{\omega} = 0.45$, at the center of the second stop gap. 
Surprisingly, there is no peak near the center of the first stop gap at variance with the $1^{st}$ corollary. 
This spectrum reveals $4$ troughs between zero frequency and the first stop gap, the $5^{th}$ trough near the center of the first stop gap and $2$ troughs between the first and second order stop gap; which seems mutually inconsistent and at variance with the $2^{nd}$ corollary. 
For p-polarized reflectivity in Fig.~\ref{fig:reflectivity-and-bandstructure} (iii), we observe a reflectivity peak near the $\tilde{\omega} = 0.45$, which corresponds to the center of the second stop gap. 
Also, no reflectivity peak appears near the center of the first stop gap at variance with the $1^{st}$ corollary. 
In this spectrum, there are $4$ troughs between zero frequency and the first stop gap and $3$ troughs between the first and second order stop gap; which seems mutually inconsistent and at variance with the $2^{nd}$ corollary. 
Therefore, the above observations for p-polarization do not agree with the observations for s-polarization. 

To remedy this seeming disagreement, we consider the geometrical structure factor $S_{\textbf{K}}$ that indicates the degree to which interference of waves scattered from identical ions within the crystal basis inside the unit cell affect the intensity of a Bragg peak associated with reciprocal lattice vector $\textbf{K}$~\cite{Ashcroft1976Book}. 
Since the intensity of the Bragg peak is proportional to the square of the absolute value of $S_{\textbf{K}}$, the Bragg peak vanishes when $S_{\textbf{K}}$ vanishes.
For a conventional cubic unit cell of the monatomic diamond structure, $S_{\textbf{K}}$ equals zero if the sum of Miller indices equals twice an odd number $n$: $h + k + l = 2n$. 
In Fig.~\ref{fig:reflectivity-and-bandstructure}, the stop gap near $\tilde{\omega} = 0.31$ in $\Gamma Z$ direction corresponds to a first-order stop gap for $hkl = \{110\}$ lattice planes in the conventional diamond structure.~\cite{Ashcroft1976Book} 
Since the sum of Miller indices in $\{110\}$ is twice the odd number $1$, the first-order stop gap in the cubic inverse woodpile photonic structure has zero geometrical structure factor and hence no zero associated Bragg reflection. 
If the sum of Miller indices $h + k + l$ is twice an even number, $S_{\textbf{K}}$ is maximum and equals two.
The stop gap near $\tilde{\omega} = 0.4$ in Fig.~\ref{fig:reflectivity-and-bandstructure} is a second-order stop gap for $hkl = \{110\}$ and corresponds to $hkl = \{220\}$ defined using X-ray diffraction in conventional diamond structure. 
Since the sum of Miller indices in $\{220\}$ equals twice an even number, the second-order stop gap has $S_{\textbf{K}} \neq 0$. 
Therefore, the second-order stop gap has a maximal structure factor. 
Hence, only the second-order stop gap in a cubic diamond-like inverse woodpile structure reveals appreciable Bragg reflection and should therefore be considered for the analysis of the observed Fabry-P{\'e}rot fringes in the reflectivity spectra.
 
The distance between lattice planes equals $d_{220}~=~c/2$ for the dominant second-order stop gap with Miller indices $hkl = \{220\}$. 
Therefore, the $L = 4c$ crystal thickness used in the computational cell in Fig.~\ref{fig:reflectivity-and-bandstructure} corresponds to a thickness $L = N d_{220} = 8 d_{220}$ in terms of a periodic layered medium (a Bragg stack).~\cite{Yariv1984Book} 
In Fig.~\ref{fig:reflectivity-and-bandstructure}, we observe the reflectivity peaks near $\tilde{\omega} = 0.45$ for s- and p-polarizations, which are at the center of the s- and p- stop gaps. 
This satisfies the first corollary for the periodic layered medium. 
Additionally, there are exactly $(N - 1) = 7$ troughs in the reflectivity spectra between zero frequency and the main stop gap corresponding to $N = 8$ lattice planes in the crystal, in agreement with the second corollary above. 
Furthermore, there are $(N - 2) = 6$ side lobes in the reflectivity spectra, again agreeing with $N = 8$ lattice planes by the third corollary. 
These three corollaries confirm that the number of Fabry-P{\'e}rot fringes in our reflectivity spectra agrees with the theory for a Bragg reflector.~\cite{Yariv1984Book} 
Moreover, this episode reminds us that it is the number of lattice planes that is fundamental in the thickness of a finite crystal, rather than the number of unit cells.

\subsection{Comparison between simulations and experiments}

\begin{table*}
\centering
\begin{tabular}{ |M{2.5cm}|M{2.5cm}|M{2.5cm}||M{2.5cm}|M{2.5cm}| }
\hline
\multicolumn{5}{|c|}{\centering Reflectivity peak for $\Gamma Z$ stop band\newline} \\
\hline \hline
&  Calc., $s$ & Exp., $\parallel$ & Calc., $p$  & Exp., $\bot$ \\ 
\hline
$\omega_{c}$ \newline ($cm^{-1}$) & $6609 \pm 25$ & $6250 \pm 32$& $6399 \pm 51$& $6400 \pm 32$\\
\hline
$\Delta \omega_{c}$ \newline ($cm^{-1}$) & $2068 \pm 25$ & $1900 \pm 32$& $1348 \pm 51$& $1000 \pm 32$\\
\hline
$\Delta \omega / \omega_{c}$  \newline $(\%)$& $31.3 \pm 0.4$& $30.4 \pm 0.5$ & $21.1 \pm 0.8$& $15.6 \pm 0.5$\\
\hline
$R_{Max}$\newline ($\%$)& $99.9 \pm 0.1$& $50 \pm 5$& $99.9 \pm 0.1$& $55 \pm 5$ \\
\hline
\end{tabular}
\caption{Comparison between numerical calculations and experimental results for the reflectivity peaks of the stop band in the $\Gamma Z$ direction. 
The s- and p-polarizations in our calculations correspond to the $\parallel$ and $\bot$ polarizations in the experiments~\cite{Huisman2011PRB}. 
We compare the central frequency ($\omega_{c}$), the frequency width ($\Delta \omega $), the relative frequency width ($\Delta \omega / \omega_{c} $) and the maximum reflectivity $R_{Max}$($\%$).} 
\label{table:CalculationVsExperiment}
\end{table*}
A recent experimental study by our group provided the signature of a 3D photonic band gap observed on silicon inverse woodpile photonic crystals.~\cite{Huisman2011PRB} 
The experimental results were collected on crystals with an extent of $L^3 = 12^{3}$ unit cells that were located on bulk silicon. 
This study discussed a number of limitations to the reflectivity, although no theoretical or numerical support was offered for these notions. 
In order to interpret this experiment, we now compare the calculated reflectivity peaks for the stop bands in the $\Gamma Z$ direction with the experimental results. 
In particular, we notably discuss the central frequency, the band width, the relative frequency width, and the maximum reflectivity, as summarized in Table~\ref{table:CalculationVsExperiment}. 

The central frequencies for p-polarization in the calculation and in the experiment agree well to within error bars. 
Using the Bragg diffraction condition~\cite{Ashcroft1976Book}, we can derive a relation for the central frequency $\omega_{c}$ such that
\begin{equation}
\frac{\omega_{c}n_{e}}{c}L = m \pi,
\label{eq:BraggLaw}
\end{equation}
where $n_{e}$ is the effective refractive index of the crystal, $L$ is the thickness of the crystal, and $m$ is the integer diffraction order. 
Since the central frequencies are nearly the same for p-polarization in the calculation and in the experiment, we deduce using Eqn.~\ref{eq:BraggLaw} that $n_{e}$ in the calculation is close to the one in the experiment. 

The central frequencies for s-polarization in calculation and in experiment differ by approximately $450$ $cm^{-1}$ more than the combined error bars. 
To remedy this inconsistency, we note that the spectrum in Ref.~\cite{Huisman2011PRB} (notably Fig. 4(d)) seems to reveal a double reflectivity peak at the $\Gamma Z$ stop band. 
Considering the strongest reflectivity peak as the indicator of the stop band, the central frequency in the experiment would be around $6600$ $cm^{-1}$, which agrees very well to the calculated one. 
Nevertheless, this hypothesis leaves open to question. 

The comparison of frequency widths and relative frequency widths in calculations and experiments exhibits small differences, which are outside the specified error bars. 
These differences can possibly be due to the dielectric permittivity $\epsilon$ or the $\frac{r}{a}$ ratio. 
In calculations, we adopt a constant dielectric permittivity $\epsilon = 12.1$ for the high index silicon backbone. In experiments, crystals were fabricated using real bulk silicon which have dispersion in $\epsilon$. 
Also, the ratio of pore radius $r$ with the lattice parameter $a$ is constant for all pores in the calculations. 
On the contrary, the fabrication process results in a distribution of $\frac{r}{a}$ for all pores in the crystal studied in the experiments, resulting in a different local volume fraction and hence variable effective refractive index across the crystal.

We observe noticeable differences between the reflectivity values in calculations and in experiments. These experiments have attributed the measured reflectivity to be limited by the finite thickness of the crystal, angle of incidence and the surface roughness. 
In reflectivity spectra shown in Fig.\ref{fig:reflectivity-and-bandstructure} and Fig.\ref{fig:transmission-vs-thickness}, we observe strong reflectivity peaks even for thin crystals. 
This asserts that the finite size is not a critical limiting factor for reflectivity. 
Fig.~\ref{fig:AngleFrequencyReflectivity}(A) and Fig.~\ref{fig:AngleFrequencyReflectivity}(B) show that the observed stop bands hardly change with angle of incidence. 
This observation supports the experimental assertion that intense reflectivity peaks measured with an objective with a large numerical aperture provide a faithful signature of the 3D photonic band gap. 
Further numerical studies are called to be performed in order to ascertain the impact of roughness of the crystal-air interface, and roughness inside the pores. 

\subsection{Back reflector for solar cells}
The efficiency of silicon photovoltaic cells critically depends on efficient ways to trap and absorb light.~\cite{Green2012ProgPhotovolt, Polman2012NatMater} 
It remains a challenge to have thin film c-Si solar cells trap a significant part of solar energy.~\cite{Joannopoulos2008Book} 
Increasing wafer thickness results in longer diffusion lengths, but increases costs. 
Traditionally, light trapping in solar cells rest on controlling light ray paths using geometrical optics, \textit{e.g.}, by scattering incident light via surface texturing and back reflection into the solar cells via a reflector. 
In practice, perfect scattering and reflection are difficult to obtain, which limits the attainable efficiency and power generation of solar cells. 
Recently, it has been shown that the light trapping approaches based on wave optics outperform all geometrical optics approaches for a certain range of frequencies.~\cite{Bermel2007OptExpress} 
One can employ specially nanodesigned structures, such as 3D photonic crystals, notably those with a complete 3D photonic band gap. 

The results in Fig.~\ref{fig:transmission-vs-thickness} reveal that a reflectivity in excess of $R > 99~\%$ (hence $T < 1~\%$) is found inside the stop band already for thin 3D silicon photonic band gap crystals, with a thickness as small as $L \geq 2c$ for s-polarization, and $L \geq 3c$ for p-polarization. 
In addition, our results in Fig.~\ref{fig:AngleFrequencyReflectivity} reveal that a 3D photonic crystal with a thickness of only $L = 4c$ reflect nearly all light within the band gap for any angle of incidence, and for both polarizations. 
Hence, our calculations support the assertion that a 3D silicon photonic crystal could serve as an efficient back reflector in a solar cell in order to enhance the efficiency. 

\section{Conclusions}
We have studied by numerical simulation the reflectivity of 3D photonic crystals with a $3D$ complete photonic band gap, to interpret recent experiments. 
We employed the finite element method to study crystals with the cubic diamond-like inverse woodpile structure, with a dielectric function similar to silicon. 
The crystals are surrounded by vacuum, and thus have a finite support, as in the experiments. 
Our calculated polarization-resolved reflectivity spectra show that the frequency ranges of the s- and p- stop bands agree very well with the corresponding stop gaps in the photonic band structure. 
We also assign bands in the band structure near these stop bands to have dominantly s- or p-character. 
From the intense reflectivity peaks, we infer that the maximum reflectivity observed in the experiments is not limited by finite size of the crystal. 
We find that the Bragg attenuation lengths in the stop bands exceed the earlier estimates based on the width of the stop band by a factor of $6$ to $9$. 
We observe that the stop band hardly changes with incident angle, which supports the experimental notion that strong reflectivity peaks measured with large numerical aperture gives faithful signature of the 3D band gap. 
In addition, we observe intriguing hybridization of the zero reflection of Fabry-P{\'e}rot fringes and the Brewster angle in our calculations, which is plausibly a characteristic property of 3D photonic band gap crystals. 
Our calculations also suggest that 3D silicon photonic band gap crystals merit study as possible candidates for back reflectors in a solar cell in order to enhance the photovoltaic efficiency.

\section{Acknowledgments}
It is a pleasure to thank Olindo Isabella, Ad Lagendijk, Oluwafemi Ojambati, Pepijn Pinkse, and Ravitej Uppu for stimulating discussions. 
WLV wishes to thank Ho-Kwang Mao, Larry Finger, and Russell Hemley for reminding him in 1991 of the crystal structure factor and its role in forbidden Bragg peaks of diamond. 
This research is supported by the Shell-NWO/FOM programme "Computational Sciences for Energy Research" (CSER). 
We also acknowledge FOM programm "Stirring of light!", as well as NWO, STW, and the MESA+ Institute for support.

\appendix
\section{Analytical validation of the numerical scheme with a semi-infinite homogeneous medium}

In order to validate our numerical scheme, we calculate reflectivity spectra of a system that can be analytically analyzed using Fresnel's equations.\cite{Griffiths1998Book} 
We consider p-polarization resolved plane waves of a single frequency with a range of angles of incidence. 
We replace the photonic crystal and the air layer on the right in Fig ~\ref{fig:ComputationalCell} (a) with a medium having dielectric permittivity $\epsilon = 12.1$, typical for silicon in the infrared regime. 
This results into a semi-infinite homogeneous medium, which is separated from the current source by an air layer. 
The finite element mesh used in the numerical calculation consists of 18732 tetrahedra per crystal unit cell (unit cell defined in terms of the lattice parameter $c$), which is less than the number of tetrahedra present in the finite element mesh used for the 3D photonic crystal.
Also, the angular resolution is $2 \degree$.

\begin{figure}
\centering
\includegraphics[width=1.0\columnwidth]{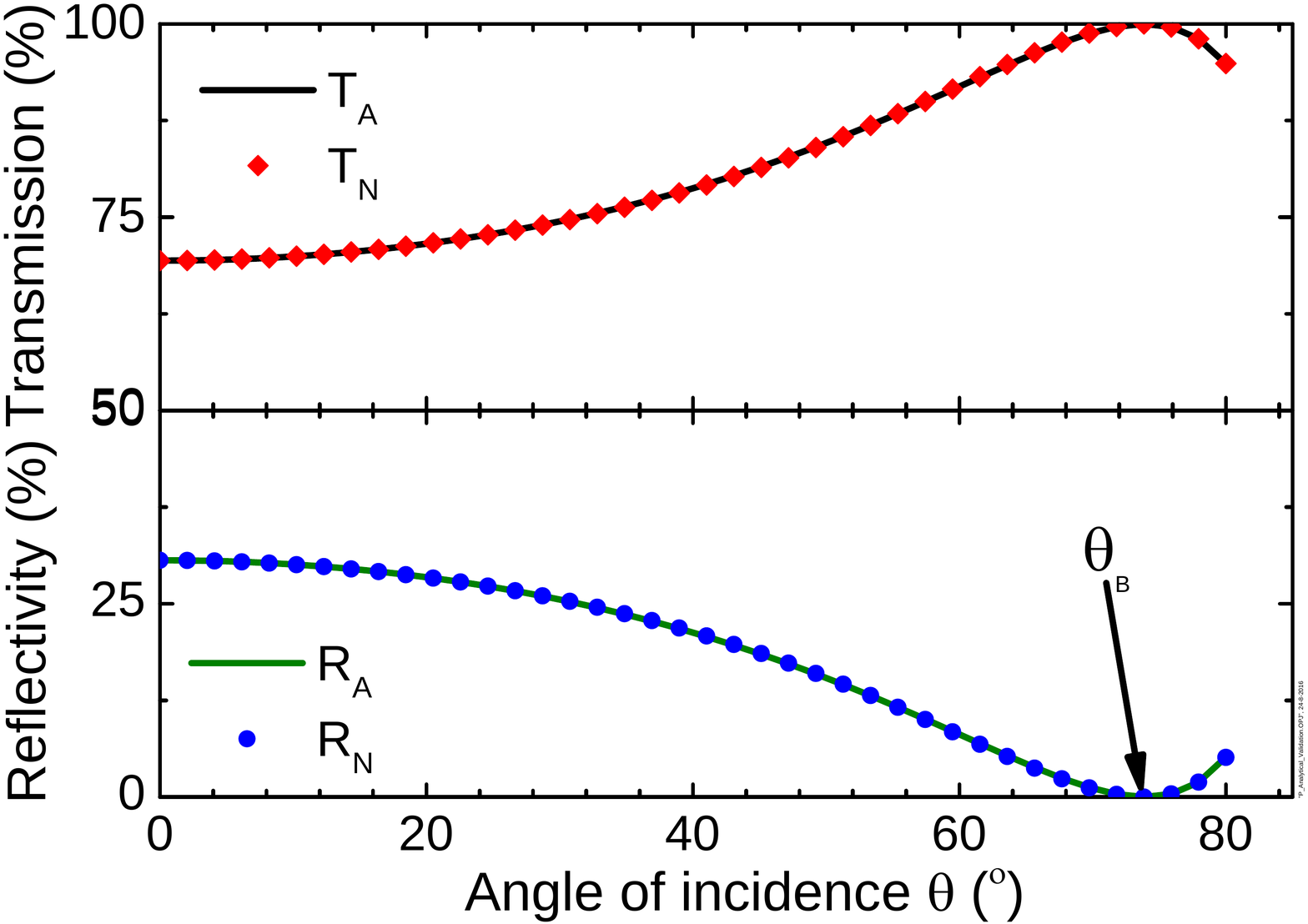}
\caption{ 
Analytical calculation versus numerical computation for reflection and transmission spectra of semi-infinite dielectric medium for p-polarization. The medium has dielectric permittivity $\epsilon = 12.1$.
The analytically calculated reflectivity ($R_{A}$) and transmission ($T_{A}$) are shown in green and black lines respectively. 
Blue circles and red diamonds represent the numerically computed reflectivity ($R_{N}$) and transmission ($T_{N}$), respectively. 
$\theta_{B}$ denotes the Brewster angle.} 
\label{fig:AnalyticalValidation_P}
\end{figure}

In Fig.~\ref{fig:AnalyticalValidation_P}, we show the calculated reflectivity and transmission spectra of a semi-infinite homogeneous medium for the above defined computational cell. 
We note that the numerical calculation agrees very well with the analytical calculation. 
We also observe the Brewster angle at $\theta_{B} = 74 \degree$, which matches the value obtained from an analytical calculation~\cite{Griffiths1998Book}. 
To calculate the relative error $\delta T_{rel}$ between the numerical calculation and the analytical result, we employ the definition

\begin{equation}
\delta T_{rel} \equiv \frac{1}{n}~\sqrt[]{\left(\sum_{i = 1}^{n} \left(\frac{(T_{A,i} - T_{N,i})^2}{T_{A,i}^2} + \frac{(R_{A,i} - R_{N,i})^2}{R_{A,i}^2}\right)\right)}
\label{eq:Error_NumericalAnalytical}
\end{equation}
with $(T_{N,i},R_{N,i})$ the numerical transmission and reflectivity, and $(T_{A,i},R_{A,i})$ the analytical transmission and reflectivity. 
For the solution shown in Fig.~\ref{fig:AnalyticalValidation_P}, the error is only about $\delta T_{rel} = 6 \times 10^{-4}$, hence we consider it to be converged. 

\begin{figure}
\centering
\includegraphics[width=1.0\columnwidth]{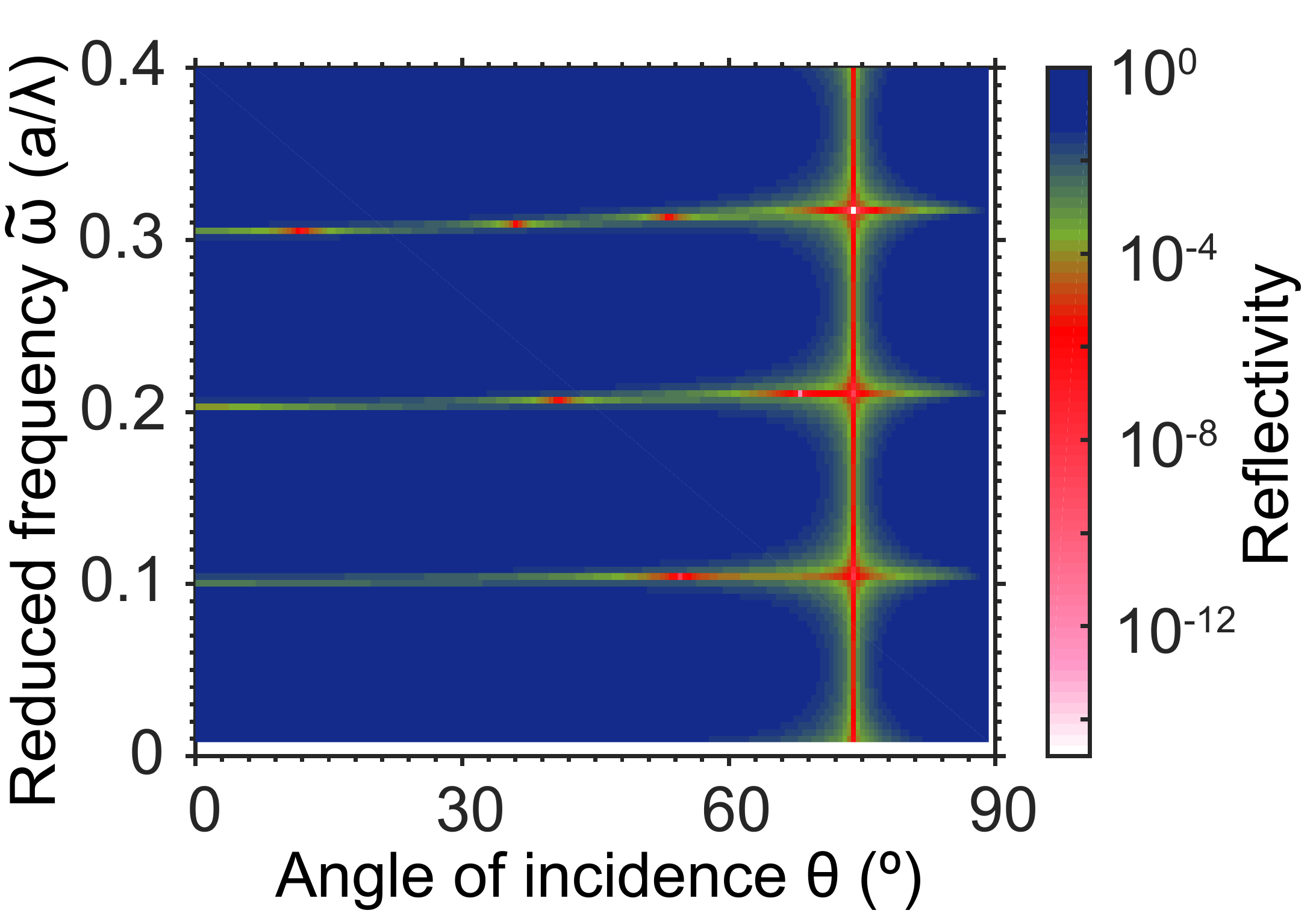}
\caption{ 
Analytically calculated angle- and frequency-resolved reflectivity spectra of a thin dielectric film for p-polarization. 
The film has a dielectric permittivity $\epsilon = 12.1$. 
The Brewster angle at $\theta_{B} = 73 \degree$ is constant with frequency. } 
\label{fig:ThinFilm_BrewsterAngle}
\end{figure}

\section{Brewster angle for a thin film}
In order to find the dependence of the Brewster angle on frequency, we analytically calculated the angle-resolved and frequency-resolved spectra for a thin film ~\cite{Ghatak1999Book}. 
We consider p-polarization resolved incident waves for angles of incidence up to $89 \degree$ off the normal. 
Figure~\ref{fig:ThinFilm_BrewsterAngle} shows Fabry-P{\'e}rot fringes corresponding to the standing waves in the thin dielectric film. 
We note that the Fabry-P{\'e}rot fringes have a nearly constant frequency for all angles of incidence.
We also observe a Brewster angle at $\theta_{B} = 73 \degree$, that is independent of frequency. 


\end{document}